\documentclass[journal,10pt,twoside]{IEEEtran}

\usepackage{latexsym}
\usepackage{amstext} 
\usepackage[centertags]{amsmath} 
\usepackage{amsfonts} 
\usepackage{amssymb}
\usepackage{mathrsfs}
\usepackage{cases}
\usepackage{array}
\usepackage{lineno} 
\usepackage{algorithm,algorithmic}
\usepackage{subfigure}
\usepackage{epsfig}
\usepackage{epstopdf}
\usepackage{graphicx}
\usepackage{psfrag}
\usepackage{multirow}

\usepackage[noadjust]{cite}
\usepackage[pdftex,hypertexnames=false,linktocpage=true]{hyperref}
\hypersetup{colorlinks=true,linkcolor=blue,anchorcolor=blue,citecolor=blue,filecolor=blue,urlcolor=blue,bookmarksnumbered=true,pdfview=FitB}
\usepackage{url}

\usepackage{blindtext}	

\usepackage{color}
\definecolor{gray}{RGB}{128,128,128}

\def\QEDclosed{\mbox{\rule[0pt]{1.3ex}{1.3ex}}} 

\def\QED{\QEDclosed} 

\usepackage{slashbox}

\usepackage{algorithm,algorithmic,xspace}
\usepackage[algo2e,vlined,linesnumbered,boxruled]{algorithm2e}

\let\oldnl\nl
\newcommand{\nonl}{\renewcommand{\nl}{\let\nl\oldnl}}

\begin{document}

\title{Transactive Energy System Deployment over Insecure Communication Links}

\author{Yang Lu, Jianming Lian, Minghui Zhu and Ke Ma 
\thanks{This work was supported partially by the Laboratory Directed Research and Development (LDRD) Program at Oak Ridge National Laboratory (ORNL) and partially by the National Science Foundation (NSF) under CAREER Grant ECCS-1846706. ORNL is operated by UT-Battelle, LLC for the U.S. Department of Energy (DOE) under Contract No. DE-AC05-00OR22725. The U.S. Government retains and the publisher, by accepting the article for publication, acknowledges that the U.S. Government retains a non-exclusive, paid-up, irrevocable, world-wide license to publish or reproduce the published form of this manuscript, or allow others to do so, for U.S. Government purposes. The DOE will provide public access to these results of federally sponsored research in accordance with the DOE Public Access Plan (http://energy.gov/downloads/doe-public-access-plan).}%
\thanks{Y. Lu is with the System Security Group, Lancaster University, Lancaster LA1 4YW, UK.}
\thanks{J. Lian is with the Grid-interactive Controls Group, Oak Ridge National Laboratory, Oak Ridge, TN, 37831, USA.}
\thanks{M. Zhu is with the School of Electrical Engineering and Computer Science, Pennsylvania State University, University Park, PA 16802, USA.}
\thanks{K. Ma is with the Optimization and Control Group, Pacific Northwest National Laboratory, Richland, WA, 99354, USA.}
\thanks{Corresponding author: Jianming Lian (e-mail: lianj@ornl.gov).}
}

\maketitle

\begin{abstract}

In this paper, the privacy and security issues associated with the transactive energy system (TES) deployment over insecure communication links are addressed. In particular, it is ensured that (1) individual agents' bidding information is kept private throughout hierarchical market-based interactions; and (2) any extraneous data injection attack can be quickly and easily detected. An implementation framework is proposed to enable the cryptography-based enhancement of privacy and security for the deployment of any general hierarchical systems including TESs. Under the proposed framework, a unified cryptography-based approach is developed to achieve both privacy and security simultaneously. Specifically, privacy preservation is realized by an enhanced Paillier encryption scheme, where a block design is proposed to significantly improve computational efficiency. Attack detection is further achieved by an enhanced Paillier digital signature scheme, where a stamp-concatenation mechanism is proposed to enable detection of data replace and reorder attacks. 
Simulation results verify the effectiveness of the proposed cyber-resilient design for transactive energy systems.
\end{abstract}


\section{Introduction}

\subsection{Background and Motivation}

Transactive control is now emerging from the electric power system as a new type of control that incorporates economic concepts and principles into the decision making and controller design of individual entities of a system. Various transactive energy system (TES) designs have been proposed to use the market clearing prices for the coordination and control of distributed energy resource (see~\cite{SL-JL-AC-WZ:2019} and the references therein). However, the market-based interactions among energy suppliers and customers inevitably raise significant concerns of privacy and security. The exchanged information on individual supply and demand curves can infer very crucial private information \cite{YG-YC-YG-YF:2016}, e.g., business secrets or personal preferences. In addition, if the communication links are insecure, the exchanged information could also be tampered by extraneous data injection attacks. Hence, the privacy and security issues necessitate the novel TES designs that can execute transactive control while simultaneously protecting data privacy and detecting malicious attacks over insecure communication links.

\subsection{Related Works}

Various techniques have been proposed in the literature to protect data privacy in power systems. In \cite{OT-DG-HVP:2013} and \cite{SH-UT-GJP:2016}, mutual information has been used to define data privacy of smart meters. This privacy metric quantifies the posterior information entropy of private data given statistical models of the source data and auxiliary information. 
In \cite{ARB-DKM-BCL-PR:2013} and \cite{ARB-DKM-PR-BCL:2012}, the technique of obfuscation has been used to protect coefficient privacy in centralized optimal power flow (OPF) problems in cloud computing. This technique masks the original OPF problem by an obfuscation transformation. Once the obfuscated problem is solved, an optimal solution to the original problem can be obtained by inverting the transformation. Differential privacy \cite{CD:06,Dwork} has been applied to the OPF \cite{ZY-PC-JC:2017,FZ-JA-SHL:2019}, economic dispatch \cite{XL-RT-DKYY-PC:2017} and thermal inertial load management \cite{7742416}. 
Differentially private schemes add random noises into individual data in such a way that they cannot be inferred by the adversaries who can access arbitrary auxiliary information. 
Our recent review paper \cite{YL-MZ:2019ARC} provides detailed comparisons of the aforementioned three techniques and homomorphic encryption (to be discussed soon) in the context of cyber-physical systems (CPSs).


On the other hand, digital signature has been widely used by the communication community for enhanced security \cite{DS-NH-GB:1995,LX-GRA:2001,TJ-YH-SZ:2004}. 
It enables the receiver to easily verify whether the digital message from the sender has been tampered or not by checking certain mathematical relations for the message and the signature. Recently, digital signature has been applied for secure communications in data aggregation in smart meters \cite{6578183}. However, the technique in \cite{6578183} cannot detect data replace or reorder attacks. Please refer to Section \ref{sec:attack detection} for details of these two attacks. For the problem of \cite{6578183}, in each cycle, each smart meter only has one data to be communicated, and the gap between two cycles could be long. Hence, these two attacks can be avoided by using a fresh new key to perform digital signature for each cycle. In contrast, detection of these two attacks is crucial for TESs. This is because, for each supplier or customer, a large number of sampled points of its supply/demand curve need to be communicated within a short period of market cycle, and it is unrealistic to adopt a fresh new key to perform digital signature for each sampled point. If the same key is used to perform digital signature for multiple sampled data, then it is possible for an attacker to launch replace and reorder attacks.

In this paper, a cyber-resilient TES design is proposed for the first time to overcome both the privacy and security issues of TESs over insecure communication links. In particular, Paillier encryption and Paillier digital signature \cite{PP:1999} are applied for the privacy-preserving and security-aware designs, respectively. Paillier encryption is an additively homomorphic encryption scheme. Homomorphic encryption is a cryptographic technique that allows algebraic operations to be carried out on ciphertexts, thus generating an encrypted result which, when decrypted, matches that of the same operations over plaintexts. It has an appealing advantage that it can achieve perfect correctness in secure multiparty computation, i.e., the computation process provides each party the correct result of its target computation without disclosing any information of its private data to the other entities. 
Homomorphic encryption has been increasingly used by the control community to achieve secure multiparty computation for optimization and control \cite{YL-MZ:2015,YL-MZ:2018Automatica,YS-KG-AA-GJP-SAS-MS-PT:2016,KK-TF:2015,FF-IS-NB:2017,NF-PP:2016,MR-HG-YW:2018}. 
In power systems, it has been adopted to data aggregation in smart meters \cite{RP:2010,FDG-BJ:2010,FL-BL-PL:2010,6578183}, and very recently in OPF problems \cite{9444341}. 
All these works adopt point-wise encryption, i.e., an encryption operation has to been done for each private data sample. This limits their usage in applications such as TESs where a large number of data samples need to be encrypted in a short period of time. Specifically, in TESs, to maintain a high market clearing accuracy, a short sampling period should be adopted and hence a large number of sampled points need to be encrypted within a market cycle. In addition, the above works on smart meters only consider integer-valued data. The work \cite{9444341} claims to be able to deal with real-valued data, but does not provide design details. Integer-valued data is enough for smart meters because smart meters readings are always integers. However, supply and demand in TESs are usually real numbers. Hence, one needs to customize standard homomorphic encryption schemes to deal with real numbers.

\subsection{Contributions}

First, privacy and security issues are identified for both hierarchical and distributed market clearing-based TESs. Then a practical framework is proposed to enable the implementation of cryptography-based approaches for general hierarchical systems including TESs. Under the proposed framework, the market participants perform Paillier encryption over the sampled points of their supply or demand curves using the coordinator's public key, and a third party is introduced to aggregate those encrypted sampled values. Then, the coordinator decrypts the aggregated encrypted sampled values using its private key. Pre- and post-operations are integrated into the encryption scheme to deal with real-valued sampled points of supply and demand curves. It is worth noting that, no participant, including the coordinator and third party, is assumed to be trustworthy. 
In this process, the coordinator has no access to individual encrypted sampled values and thus cannot recover individual supply or demand curves. Without knowing the coordinator's private key, the third party and the eavesdroppers over insecure communication links, cannot recover individual supply or demand curves either. Preliminary results on the privacy-preserving TES design was presented in \cite{YL-JL-MZ:2020}.

The privacy-preserving TES design by directly integrating Paillier encryption is not ready for practical implementation yet. First, the associated computational overhead is proportional to the number of sampled points. When this number becomes large, the process of encryption and decryption would be time-consuming and may not be suitable for real-time market operations. To address this computational issue, a block design is further proposed in this paper to improve the computational efficiency by the number of sampled points times while still maintaining the level of privacy.  
Second, the security issue has not been addressed in the presence of potential data injection attacks over insecure communication links. In this paper, an attack detection mechanism based on Paillier digital signature is proposed. When sending the data over the insecure communication link, the sender first generates a digital signature for the data using its own private key and then sends the data together with its signature to the receiver. After receiving the data, the receiver can perform a verification operation using the sender's public key to detect whether the received data and signature has been tampered or not. Without knowing the sender's private key, an attacker is not able to generate a pair that can pass the receiver's verification. Specifically, to detect data replace and reorder attacks, we customize the standard Paillier digital signature scheme by concatenating a stamp to each message to identify its unique index, and a digital signature is generated for the stamped message. With this mechanism, the data replace or reorder attacks can no longer pass the verification operation as a replaced or reordered pair of message and signature does not match the index.
%
The efficacy of the overall proposed cyber-resilient TES design is verified via simulation results.

A preliminary version of this paper is presented in \cite{YL-JL-MZ:2020}. Compared with \cite{YL-JL-MZ:2020}, the current paper includes data injection attack and proposes a security-aware mechanism, and develops a block design that can improve computational efficiency.

\subsection{Organization}

The rest of this paper is organized as follows. In Section \ref{sec:problem statement}, the TES is briefly introduced with the privacy and security issues identified. In Section \ref{sec:privacy design}, a privacy-preserving TES design is first developed based on the Paillier encryption scheme, and then an attack detection algorithm based on the Paillier digital signature is proposed in Section \ref{sec:address security issue} for security-aware TES design. In Section \ref{sec:simulation}, case studies are presented to illustrate the effectiveness of the proposed 
cyber-resilient TES design. Conclusions are found in Section~\ref{sec:conclusion}.

\subsection{Notations}

Denote by $\mathbb{R}$ and $\mathbb{N}$ the sets of real and natural numbers (including 0), respectively. Given a positive integer $n$, let $\mathbb{Z}_n=\{0,1,\cdots,n-1\}$ and let $\mathbb{Z}_n^*$ denote the set of positive integers that are smaller than and co-prime to $n$. Given positive integers $x$ and $y$, denote by ${\rm gcd}(x,y)$ and ${\rm lcm}(x,y)$ the greatest common divisor and the least common multiple of $x$ and $y$, respectively. 
Given $x,y\in\mathbb{N}$, denote by $x\leftrightarrow y$ the concatenation of $x$ and $y$, e.g., $12\leftrightarrow 345=12345$. Given $x\in\mathbb{N}$, denote by ${\rm num}(x)$ the number of digits in $x$, e.g., ${\rm num}(123)=3$. Given $x\in\mathbb{N}$ and two positive integers $a\leq b\leq{\rm num}(x)$, denote by $[x]_{a:b}$ the part from the $a$-th digit to the $b$-th digit of $x$, with the first digit being the leftmost one, and denote by $[x]_{a:{\rm end}}$ the part from the $a$-th digit to the last digit of $x$, e.g., $[12345]_{2:4}=234$ and $[12345]_{3:{\rm end}}=345$. Given $a\in\mathbb{N}$, denote by $(0\cdots0)_a$ the concatenation of $a$ zeros, e.g., $(0\cdots0)_3=000$.

\section{Problem Statement}\label{sec:problem statement}

In this section, we first briefly introduce the concept of TES. Then, we assess the cyber vulnerabilities of TES to identify the privacy and security issues associated with the existing TES designs. Finally, we state the objective of this paper.

\subsection{Transactive Energy System\label{sec:TES}}

\begin{figure}
\begin{center}
\includegraphics[width=0.75\linewidth]{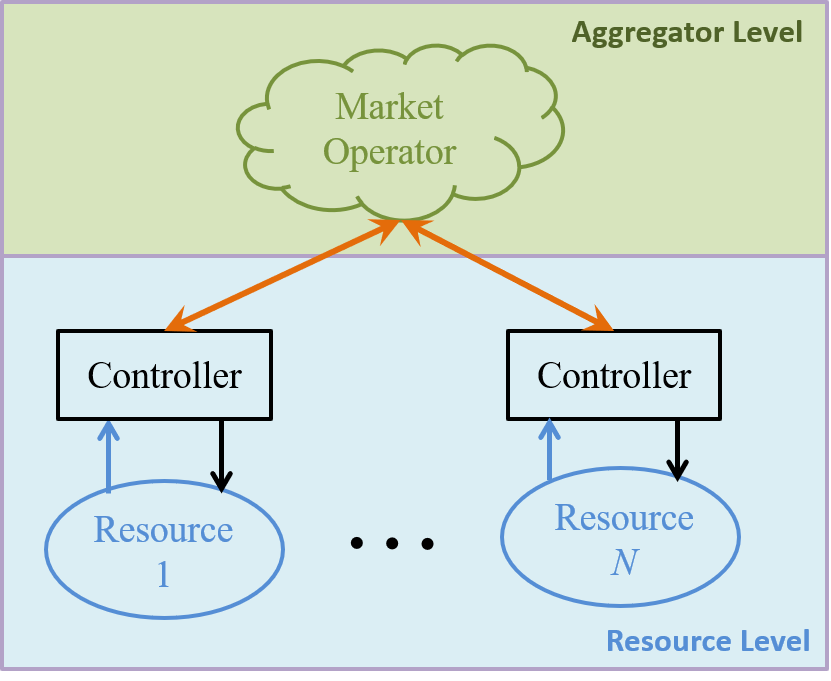}
\caption{Illustration of the underlying hierarchical structure of TESs.}
\label{TES_structure}
\end{center}
\end{figure}

Within the TES, different entities can be classified into three types: coordinator (CO), supplier, and customer, where the coordinator is the market operator, a supplier is an energy seller, and a customer is an energy buyer. 
The TES can then be modeled as a multi-agent system with a hierarchical structure, as shown in Fig. \ref{TES_structure}. 
The coordinator aims to allocate energy resources to the suppliers and customers to ensure both individual and social objectives and constraints. This is referred to as the resource coordination problem. In transactive coordination, the coordinator achieves the optimal resource coordination by properly setting the resource price, which is called the market clearing price. We next present a typical TES to illustrate transactive coordination.

Denote by $\mathcal{V}_s$ and $\mathcal{V}_d$ be the set of suppliers and the set of customers, respectively. 
Let $N_s=|\mathcal{V}_s|$, $N_d=|\mathcal{V}_d|$, and $\mathcal{V}\triangleq\mathcal{V}_s\cup\mathcal{V}_d$. In the remaining of the paper, when it is necessary to differentiate between suppliers and customers, we will use ``supplier $i\in\mathcal{V}_s$'' or ``customer $i\in\mathcal{V}_d$''. Otherwise, we will use ``agent $i\in\mathcal{V}$''.

Given a market clearing price $\lambda$, each supplier $i\in\mathcal{V}_s$ aims to find an optimal supply that maximizes its profit, defined as the earning in energy selling minus the cost in energy generation. The profit optimization problem of supplier $i\in\mathcal{V}_s$ is formulated as
\begin{align*}
    \max_{p_i^s\in\mathcal{L}_i^s}\lambda p_i^s-C_i(p_i^s)
\end{align*}
where $p_i^s$ is its supply, $C_i:\mathbb{R}\to\mathbb{R}$ is its cost function,
$\lambda$ is the resource price, and $\mathcal{L}_i^s$ is the feasible set of $p_i^s$.

Given a resource price $\lambda$, each customer $i\in\mathcal{V}_d$ aims to find an optimal demand that maximizes its utility, defined as the benefit in energy usage minus the cost in energy purchasing. The utility optimization problem of customer $i\in\mathcal{V}_d$ is formulated as
\begin{align*}
    \max_{p_i^d\in\mathcal{L}_i^d}U_i(p_i^d)-\lambda p_i^d
\end{align*}
where $p_i^d$ is its demand, $U_i:\mathbb{R}\to\mathbb{R}$ is its utility function,
and $\mathcal{L}_i^d$ is the feasible set of $p_i^d$.

The suppliers and customers take best response with respect to the resource price given by the coordinator. The coordinator then aims to select a resource price that maximizes the social welfare. The bi-level optimization problem of the coordinator is formulated as
\begin{IEEEeqnarray}{r'l}
\label{OP coordinator}
    \max_{\lambda\in\mathbb{R}} & \sum_{i\in\mathcal{V}_d}U_i\left(p_i^{d*}(\lambda)\right)-\sum_{i\in\mathcal{V}_s}C_i\left(p_i^{s*}(\lambda)\right)
    \IEEEyesnumber \IEEEyessubnumber \label{OP CO objective}\\
{\rm s.t.} & p_i^{s*}(\lambda)=\mathop{\rm{argmax}}\limits_{p_i^s\in\mathcal{L}_i^s} \lambda p_i^s-C_i(p_i^s),\;\forall i\in\mathcal{V}_s,
\IEEEyessubnumber \label{OP CO constraint 1}\\
& p_i^{d*}(\lambda)=\mathop{\rm{argmax}}\limits_{p_i^d\in\mathcal{L}_i^d} U_i(p_i^d)-\lambda p_i^d,\;\forall i\in\mathcal{V}_d,
\IEEEyessubnumber \label{OP CO constraint 2}\\
& \sum_{i\in\mathcal{V}_s}p_i^{s*}(\lambda)=\sum_{i\in\mathcal{V}_d}p_i^{d*}(\lambda).
\IEEEyessubnumber \label{OP CO constraint 3}
\end{IEEEeqnarray}

In problem \eqref{OP coordinator}, the function $p_i^{s*}$ (resp. $p_i^{d*}$) is called the supply (resp. demand) function, and its graphical representation is called the supply (resp. demand) curve. 
Both hierarchical and distributed market clearing approaches have been widely used to determine the optimal solution $\lambda^*$ of problem \eqref{OP coordinator}. We next briefly discuss these two approaches.

\textit{Hierarchical market clearing.}
Hierarchical market clearing is implemented through auction. Individual suppliers $i\in\mathcal{V}_s$ and customers $i\in\mathcal{V}_d$ submit their entire supply or demand curves to the coordinator, respectively. Upon receiving all the individual curves, the coordinator first determines the aggregated supply and demand curves, and then find the market clearing price as the intersection between the aggregated supply and demand curves.


\textit{Distributed market clearing.} Distributed market clearing works in an iterative manner. At the $k$-th iteration, the coordinator broadcasts an estimated price $\lambda(k)$ to the market participants. Individual suppliers $i\in\mathcal{V}_s$ and customers $i\in\mathcal{V}_d$ determine $p_i^s(k)=p_i^{s*}(\lambda(k))$ and $p_i^d(k)=p_i^{d*}(\lambda(k))$, respectively, and report them to the coordinator. Then the coordinator updates the price estimate for the next iteration until the market clearing price converges.

\subsection{Cyber Vulnerabilities\label{sec:privacy and security issues}}

TES is in fact a very typical CPS, where the market clearing prices are determined in the cyber space and the control tasks are performed in the physical world. Hence, TESs share the typical cyber vulnerabilities of general CPSs. In this paper, we consider the privacy and security issues associated with TESs. This subsection identifies these issues with respect to the two market clearing approaches introduced above.

\subsubsection{Privacy issue\label{sec:privacy issue}}

The hierarchical market clearing requires individual agents to submit their supply or demand curves to the coordinator. With this information, the coordinator or an eavesdropper over the insecure communication links can easily infer individual cost or utility functions. 
In fact, the inverse supply or demand function is just the derivative of the corresponding cost or utility function \cite{JL-HR-YS-DJH:2019}. 
Hence, individual cost or utility functions can be recovered by integrating the inverse of the corresponding supply or demand functions. 
This could potentially expose the business secrets (for suppliers) or personal preferences (for customers). We refer to the problem of private data leakage as the privacy issue.

The distributed market clearing can partially mitigate the privacy issue as individual agents do not submit their supply or demand curves to the coordinator, but only those quantities with respect to the broadcasted prices.
However, 
the coordinator could make use of the iterative nature of the distributed approach to intentionally broadcast a large number of prices covering the entire admissible range. In this way, the coordinator or an eavesdropper could still recover individual supply or demand curves arbitrarily well.

\subsubsection{Security issue\label{sec:security issue}}

Both market clearing approaches require information exchange between the coordinator and the agents. If the communication links are unauthenticated, extraneous attackers can send forged information to legitimate participants or tamper the information in transit to disrupt the market operation. This is termed as data injection attack (also known as data integrity attack or data tampering attack). In the presence of such attacks, the data received by the coordinator could be completely distorted, and the clearing price determined accordingly could arbitrarily deviate from the true clearing price and may lead to market chaos. We refer to the problem of data forging and tampering as the security issue.


\subsection{Objectives\label{sec:objective}}

In this paper, we aim to develop a cyber-resilient TES design that simultaneously satisfies the following three properties:

(1) Correctness: The coordinator can determine the correct clearing price $\lambda^*$ such that $\sum_{i\in\mathcal{V}_s}p_i^{s*}(\lambda^*)=\sum_{i\in\mathcal{V}_d}p_i^{d*}(\lambda^*)$;

(2) Privacy preservation: After the execution of the algorithm, for each supplier $i\in\mathcal{V}_s$ (resp. customer $i\in\mathcal{V}_d$), no other entity can infer the value of $p_i^{s*}(\lambda)$ (resp. $p_i^{d*}(\lambda)$) for any admissible $\lambda$;

(3) Security awareness: Any extraneous data injection attacks can be detected by legitimate message receivers.

For the purpose of illustration, only the hierarchical market clearing is considered in the following. However, the proposed design can be easily extended to distributed market clearing.

\section{Privacy-preserving design\label{sec:privacy design}}

In this section, the privacy-preserving TES design is developed based on homomorphic encryption. We first propose a framework for practical deployment. Then we define the attacker model adopted in this section. After that, we present the details of the proposed privacy-preserving design. Finally, we propose an approach to ensure the computational efficiency for practical implementation.

\subsection{Implementation Framework}

\begin{figure}
\begin{center}
\includegraphics[width=0.85\linewidth]{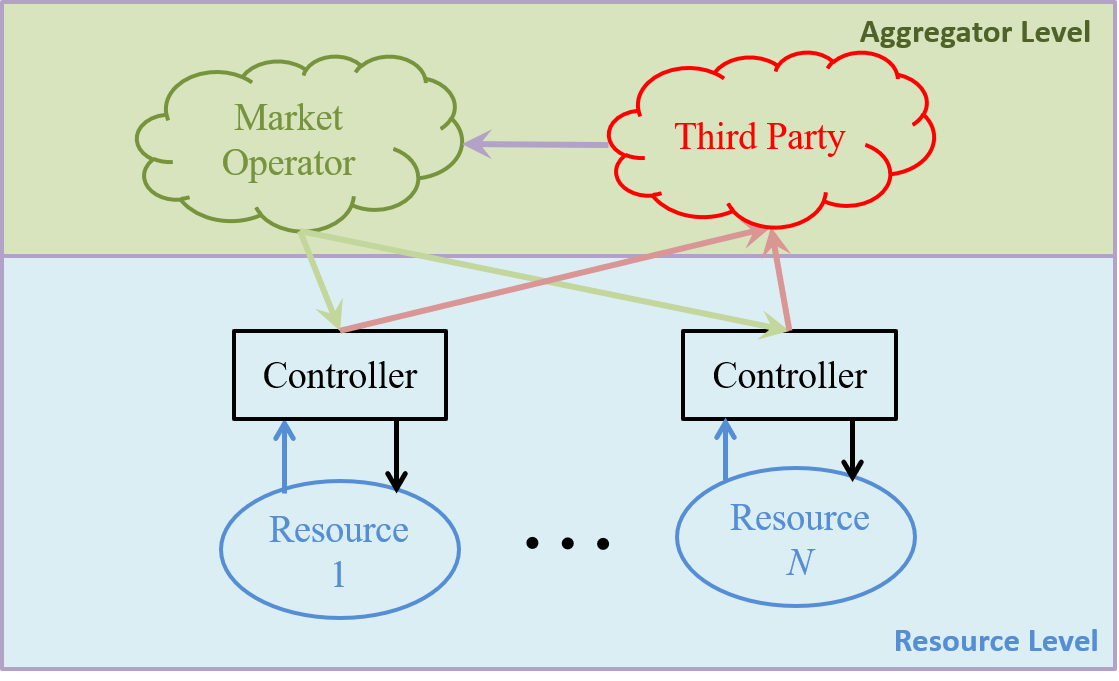}
\caption{Illustration of the proposed framework for TES deployment.}
\label{TES_aug_structure}
\end{center}
\end{figure}

In order to preserve the privacy, it requires that the coordinator should obtain the aggregated curve without knowing individual ones. In cryptography, homomorpihc encryption is a promising technique to fulfill this requirement. This technique requires that the entity who receives individual ciphertexts and carries out algebraic operations to be different from the entity who performs the decryptions. Hence, in order to enable the use of homomorphic encryption, we introduce an additional third party (TP) as the independent entity who is responsible of receiving individual ciphertexts and performing encrypted aggregations. 
The proposed framework is shown in Fig. \ref{TES_aug_structure}, in which 
we assume that there is a communication link $(i,{\rm TP})$ between each agent $i\in\mathcal{V}$ and the third party, and a communication link $({\rm TP},{\rm CO})$ between the third party and the coordinator. The third party can be implemented by an extraneous entity, e.g., a cloud service provider. Indeed, the third-party cloud service, for example, the IBM Power Virtual Server \cite{IBMcloud}, has emerged in power systems to support thosee applications that are computationally intensive.

\subsection{Attacker Model\label{sec:attacker model and security notion}}



We assume that any market participant $i\in\mathcal{V}\cup\{{\rm CO},{\rm TP}\}$ is semi-honest, i.e., it correctly follows the designed algorithm but attempts to use received messages  
to infer other participants' private data (\cite{Hazay}, pp-20). 
In addition, there could be external attackers that can eavesdrop the communication links. 
In this section, we assume that there are no data injection attacks. Such attacks are considered in the next section.

\subsection{Algorithm Design\label{sec:algorithm design}}

This subsection presents the proposed privacy-preserving auction-based algorithm. 
In plain auction-based clearing in Section \ref{sec:TES}, individual supply or demand curves are sampled and discrete-valued versions are submitted. Denote by $\lambda_{\min}$ and $\lambda_{\max}$ the lower and upper bounds of resource price, respectively. Denote by $\tau$ the sampling period and $N_p$ the number of sampled values. 
For each supplier $i\in\mathcal{V}_s$ (resp. customer $i\in\mathcal{V}_d$), denote by $p_{i\ell}^{s*}$ (resp. $p_{i\ell}^{d*}$) its $\ell$-th sampled value, i.e., $p_{i\ell}^{s*}=p_{i}^{s*}(\lambda_{\min}+\ell\tau)$ (resp. $p_{i\ell}^{d*}=p_{i}^{d*}(\lambda_{\min}+\ell\tau)$). Denote by $\sigma\in\mathbb{N}$ the precision level of the sampled values, i.e., for any $p_{i\ell}^{s*}$ and $p_{i\ell}^{d*}$, only the first $\sigma$ decimal fraction digits are kept, while the rest are dropped. Assume that the coordinator and all the suppliers (resp. customers) know a strict upper bound $\delta_s$ (resp. $\delta_d$) of individual supply (resp. demand) curves, i.e., $\delta_s>p_i^s$ for all $i\in\mathcal{V}_s$ and all $p_i^s\in\mathcal{L}_i^s$ (resp. $\delta_d>p_i^d$ for all $i\in\mathcal{V}_d$ and all $p_i^d\in\mathcal{L}_i^d$).

Our privacy-preserving auction-based design, Algorithm \ref{algo:privacy preserving TES}, is based on the Paillier encryption scheme. Preliminaries of Paillier encryption, including the sub-algorithms ${\rm Alg_{key}}$, ${\rm Alg_{enc}}$, and ${\rm Alg_{dec}}$, are given in Appendix \ref{sec:Paillier encryption}.

\begin{algorithm2e}[htbp]
\caption{Privacy-preserving auction}\label{algo:privacy preserving TES}

\textbf{Key generation}

\nonl The CO runs $(\alpha,\beta,\nu,\pi)={\rm Alg_{key}}(n)$ such that $\alpha>\max\{10^\sigma N_s\delta_s,10^\sigma N_d\delta_d\}$, broadcasts $(\alpha,\beta)$ and keeps $(\nu,\pi)$ private to itself;

\nonl \For{$\ell=1$; $\ell\leq N_p$; $\ell=\ell+1$}{
\textbf{Encryption}

\nonl Each supplier $i\in\mathcal{V}_s$ runs
\begin{align*}
    y_{i\ell}^s={\rm Alg_{enc}} (\alpha,\beta,10^\sigma p_{i\ell}^{s*})
\end{align*}
and sends $y_{i\ell}^s$ to the TP;

\nonl Each customer $i\in\mathcal{V}_d$ runs
\begin{align*}
    y_{i\ell}^d={\rm Alg_{enc}} (\alpha,\beta,10^\sigma p_{i\ell}^{d*})
\end{align*}
and sends $y_{i\ell}^d$ to the TP;

\textbf{Computation over ciphertexts}

\nonl The TP computes
\begin{align*}
    &y_{\ell}^s=\prod_{i\in\mathcal{V}_s}y_{i\ell}^s\mod\alpha^2,\\
    &y_{\ell}^d=\prod_{i\in\mathcal{V}_d}y_{i\ell}^d\mod\alpha^2
\end{align*}
and and sends $(y_{\ell}^s,y_{\ell}^d)$ to the CO;

\textbf{Decryption}

\nonl The CO runs
\begin{align*}
    &\hat y_{\ell}^s={\rm Alg_{dec}}(\alpha,\nu,\pi,y_{\ell}^s)/10^\sigma,\\
    &\hat y_{\ell}^d={\rm Alg_{dec}}(\alpha,\nu,\pi,y_{\ell}^d)/10^\sigma;
\end{align*}
}
\textbf{Setting clearing price}

\nonl The CO sets $\lambda^*=\lambda_{\min}+\ell\tau$ such that $\hat y_{\ell}^s=\hat y_{\ell}^d$, and sends $\lambda^*$ to each agent $i\in\mathcal{V}$.
%
\end{algorithm2e}

At step 1, the coordinator generates a set of keys by the Paillier key-generation algorithm. The public keys are broadcasted while the private keys are kept private to itself. The bound on $\alpha$ is to guarantee decryption correctness. Roughly speaking, to ensure decryption correctness, the public key $\alpha$ must be larger than the computing result. Please refer to the statement of homomorphic property at the end of Appendix \ref{sec:Paillier encryption}, in which it requires $\alpha>\sum_{\ell=1}^mpt_\ell$. In our problem, $\max\{10^\sigma N_s\delta_s,10^\sigma N_d\delta_d\}$ is a strict upper bound for all computing results, i.e., sampled values of aggregated supply and demand curves. Hence, the bound on $\alpha$ guarantees decryption correctness for all computing results at step 4. Actually, for the sake of privacy, $\alpha$ needs to be very large, e.g., in the magnitude of $2^{2000}$ \cite{DG:2017}. Hence, the upper bound condition on $\alpha$ is usually automatically satisfied even if the participants do not know $\delta_s$ or $\delta_d$. At step 2, each supplier $i\in\mathcal{V}_s$ (resp. customer $i\in\mathcal{V}_d$) encrypts its sampled value $10^\sigma p_{i\ell}^{s*}$ (resp. $10^\sigma p_{i\ell}^{d*}$) by the Paillier encryption algorithm with the public keys $(\alpha,\beta)$, and sends the ciphertext $y_{i\ell}^s$ (resp. $y_{i\ell}^d$) to the third party. Notice that $10^\sigma p_{i\ell}^{s*}$ and $10^\sigma p_{i\ell}^{d*}$ are both non-negative integers. At step 3, the third party performs computations over received ciphertexts according to the homomorphic property of the Paillier encryption scheme, i.e., multiplication of ciphertexts provides an encryption of sum of plaintexts. Hence, $y_{\ell}^s$ and $y_{\ell}^d$ are actually encryptions of the $\ell$-th sampled values of the aggregated supply and demand curves, respectively. The third party then sends $y_{\ell}^s$ and $y_{\ell}^d$ to the coordinator. At step 4, the coordinator decrypts $y_{\ell}^s$ and $y_{\ell}^d$ by the Paillier decryption algorithm with its public key $\alpha$ and private keys $(\nu,\pi)$, and transforms the decrypted results back to real numbers via dividing them by $10^\sigma$. At step 5, the coordinator sets and broadcasts the clearing price $\lambda^*$.



Algorithm \ref{algo:privacy preserving TES} has the following properties:

(1) Correctness: For each $\ell\in\{1,\cdots,N_p\}$, it follows that $\hat y_{\ell}^s=\sum_{i\in\mathcal{V}_s}p_{i}^{s*}(\lambda_{\min}+\ell\tau)$ and $\hat y_{\ell}^d=\sum_{i\in\mathcal{V}_d}p_{i}^{d*}(\lambda_{\min}+\ell\tau)$.

The correctness property states that $\hat y_{\ell}^s$ and $\hat y_{\ell}^d$ are just the $\ell$-th sampled values of the original aggregated supply and demand curves, respectively. This property directly follows from the homomorphic property of the Paillier encryption scheme (please refer to the end of Appendix \ref{sec:Paillier encryption}). Since $\lambda^*$ is set as $\lambda^*=\lambda_{\min}+\ell\tau$ such that $\hat y_\ell^s=\hat y_\ell^d$, the correctness property leads to $\sum_{i\in\mathcal{V}_s}p_{i}^{s*}(\lambda^*)=\sum_{i\in\mathcal{V}_d}p_{i}^{d*}(\lambda^*)$. Hence, optimal market-based coordination is achieved.

(2) Privacy preservation: If the DCRA holds, then, after the execution of the algorithm, for each supplier $i\in\mathcal{V}_s$ (resp. customer $i\in\mathcal{V}_d$), for all $\ell\in\{1,\cdots,N_p\}$, the value of $p_i^{s*}(\lambda_{\min}+\ell\tau)$ (resp. $p_i^{d*}(\lambda_{\min}+\ell\tau)$) is semantically secure.

The privacy preservation property directly follows from the semantic security of the Paillier encryption scheme (please refer to the end of Appendix \ref{sec:Paillier encryption}). Specifically, after the execution of Algorithm \ref{algo:privacy preserving TES}, each agent $i\in\mathcal{V}$ only knows its own supply or demand curve and the market clearing price; the coordinator only knows the aggregated supply and demand curves and the market clearing price; the third party or an extraneous eavesdropper only knows the market clearing price. Therefore, any agent's individual supply or demand curve is not known to any other entity and privacy preservation is achieved.

\subsection{Block Design for Improved Computational Efficiency\label{sec:block design}}

Algorithm \ref{algo:privacy preserving TES} works in a point-wise manner, i.e., all the cryptographic operations are performed for each sampled value of the supply or demand curves. Specifically, each agent $i$ performs $N_p$ times encryption, the third party performs $2N_p$ times computation over ciphertexts, and the coordinator performs $2N_p$ times decryption. When $N_p$ is large, the implementation of Algorithm \ref{algo:privacy preserving TES} would be time-consuming. In this subsection, we propose a design such that all the cryptographic operations are performed in a block-wise manner and the numbers of the operations are independent of $N_p$.

Roughly speaking, each agent concatenates all its $N_p$ sampled values to form a single block, and all the cryptographic operations are performed over the block. To guarantee correctness, we propose to pad enough zeros in each sampled value before concatenation. In this way, each agent performs once encryption, the third party performs twice computation over ciphertexts, and the coordinator performs twice decryption. Hence, under the same key length, the computational complexity reduces approximately by $N_p$ times. 
We next detail the block design and illustrate how to embed it into Algorithm \ref{algo:privacy preserving TES}.

At step 1, change the bound of $\alpha$ as
\begin{IEEEeqnarray*}{r'l}
&\alpha>\max\left\{\frac{10^{N_p\times{\rm num}(10^\sigma N_s\delta_s)}-1}{10^{{\rm num}(10^\sigma N_s\delta_s)}-1}10^\sigma N_s\delta_s,\right. \\
&\qquad\qquad\qquad\qquad\left.\frac{10^{N_p\times{\rm num}(10^\sigma N_d\delta_d)}-1}{10^{{\rm num}(10^\sigma N_d\delta_d)}-1}10^\sigma N_d\delta_d\right\}.
\end{IEEEeqnarray*}

Before step 2, individual suppliers and customers pad their $p_{i\ell}^{s*}$ and $p_{i\ell}^{d*}$ as
\begin{align*}
    &\bar p_{i\ell}^{s*}=(0\cdots 0)_{{\rm num}(10^\sigma N_s\delta_s)-{\rm num}(10^\sigma p_{i\ell}^{s*})}\leftrightarrow 10^\sigma p_{i\ell}^{s*},\\
    &\bar p_{i\ell}^{d*}=(0\cdots 0)_{{\rm num}(10^\sigma N_d\delta_d)-{\rm num}(10^\sigma p_{i\ell}^{d*})}\leftrightarrow 10^\sigma p_{i\ell}^{d*}
\end{align*}
respectively, and then form the concatenation
\begin{align*}
    &\bar p_{i}^{s*}=\bar p_{iN_p}^{s*}\leftrightarrow \cdots\leftrightarrow \bar p_{i1}^{s*},\\
    &\bar p_{i}^{d*}=\bar p_{iN_p}^{d*}\leftrightarrow \cdots\leftrightarrow \bar p_{i1}^{d*}
\end{align*}
respectively. After that, remove the loop over sampled values (i.e., remove the $\ell$ loop and drop $\ell$ from anywhere at steps 2--4). At step 2, replace $10^\sigma p_{i\ell}^{s*}$ and $10^\sigma p_{i\ell}^{d*}$ with $\bar p_{i}^{s*}$ and $\bar p_{i}^{d*}$, respectively. After step 4, the coordinator performs an additional cutting step by setting, for each $\ell=1,\cdots,N_p$:
\begin{align*}
    &\hat y_{\ell}^s=[\hat y^s]_{({\rm end}-\ell\times {\rm num}(10^\sigma N_s\delta_s)+1):({\rm end}-(\ell-1)\times{\rm num}(10^\sigma N_s\delta_s))},\\
    &\hat y_{\ell}^d=[\hat y^d]_{({\rm end}-\ell\times {\rm num}(10^\sigma N_d\delta_d)+1):({\rm end}-(\ell-1)\times{\rm num}(10^\sigma N_d\delta_d))}.
\end{align*}

\section{Security-aware design\label{sec:address security issue}}

In this section, the security-aware TES design is proposed based on digital signature. We first extend the attacker model considered in Section~\ref{sec:privacy design} to include data injection attacks.
Then, we equip Algorithm \ref{algo:privacy preserving TES} with an attack detection mechanism to further achieve objective (3) stated in Section \ref{sec:objective}.

\subsection{Attacker Model}

In this section, all the market participants $\mathcal{V}\cup\{{\rm CO},{\rm TP}\}$ have the same attacker model described in Section \ref{sec:attacker model and security notion}. In addition, there could exist extraneous attackers that launch data injection attacks. In particular, they can send arbitrarily forged information to legitimate participants or arbitrarily tamper the information in transit. In this paper, we focus on data injection attacks over communication links, but do not consider Byzantine attacks, in which some legitimate participants arbitrarily deviate from the given algorithm. We leave the study of Byzantine attacks to our future works.

\subsection{Algorithm Design\label{sec:attack detection}}

The security-aware design, Algorithm \ref{algo:attack detection}, is based on the Paillier digital signature scheme. Preliminaries of Paillier digital signature, including  the sub-algorithms ${\rm Alg_{sig}}$ and ${\rm Alg_{ver}}$, are given in Appendix \ref{sec:Paillier digital signature}.

Consider the case where participant $i$ aims to send a message $m$ to participant $j$ via link $(i,j)$. Participant $i$ generates Paillier keys $(\alpha_i,\beta_i,\nu_i,\pi_i)$, where $(\alpha_i,\beta_i)$ are sent over an authenticated link to participant $j$ and $(\nu_i,\pi_i)$ are kept private to itself. Participants $i$ and $j$ perform an attack detection mechanism given by Algorithm \ref{algo:attack detection}. The inputs include the identity indicators $i$ and $j$, participant $i$'s keys $(\alpha_i,\beta_i,\nu_i,\pi_i)$, message $m$, and an index $\ell\in\mathbb{N}$. The outputs include a binary attack indicator ${\rm FLAG}$ and participant $j$'s output message $\bar m$. In particular, the index $\ell$ is a stamp to identify which data $m$ is. An example of setting $\ell$ is given later.

\begin{algorithm2e}[htbp]
\caption{Attack detection mechanism}\label{algo:attack detection}

\nonl Syntax: $({\rm FLAG},\bar m)={\rm Alg_{ad}}(i,j,\alpha_i,\beta_i,\nu_i,\pi_i,m,\ell)$.

\textbf{Signature}

\nonl Participant $i$ runs
\begin{align*}
    (s_1,s_2)={\rm Alg_{sig}}(\alpha_i,\beta_i,\nu_i,\pi_i,\ell\leftrightarrow m)
\end{align*}
and sends $(\ell\leftrightarrow m,s_1,s_2)$ to participant $j$;

\textbf{Verification}

\nonl On receiving the $\ell$-th triple $(\bar z,\bar s_1,\bar s_2)$ from $(i,j)$, participant $j$ sets $\bar m=[\bar z]_{{\rm num}(\ell)+1:{\rm end}}$ and ${\rm FLAG}$ $=1$ if ${\rm Alg_{ver}}(\alpha_i,\beta_i,\bar z,\bar s_1,\bar s_2)=1$ and $[\bar z]_{1:{\rm num}(\ell)}=\ell$, and sets $\bar m={\rm NULL}$ and ${\rm FLAG}=0$ otherwise.
\end{algorithm2e}

First, participant $i$ generates a pair of signatures $(s_1,s_2)$ for $\ell\leftrightarrow m$ by the Paillier signature algorithm and sends the triple $(\ell\leftrightarrow m,s_1,s_2)$ to participant $j$. Upon receiving the $\ell$-th triple $(\bar z,\bar s_1,\bar s_2)$ from $(i,j)$, participant $j$ performs a verification operation to detect whether the triple has been attacked. The triple passes the verification if and only if: (1) the triple $(\bar z,\bar s_1,\bar s_2)$ passes the Paillier verification algorithm, and (2) the index matches, i.e., the first ${\rm num}(\ell)$ digits of $\bar z$ matches $\ell$. If the triple passes the verification, then participant $j$ sets ${\rm FLAG}=1$ to indicate no attack and sets $\bar m=[\bar z]_{{\rm num}(\ell)+1:{\rm end}}$, which is just $m$. Otherwise, participant $j$ sets ${\rm FLAG}=0$ to indicate attack and sets $\bar m={\rm NULL}$. The detection is enabled by the property that, without knowing participant $i$'s private keys $(\nu_i,\pi_i)$, an attacker cannot generate a triple that can pass participant $j$'s verification. The index $\ell$ serves as the time stamp of message $m$. Without using the time stamp, a verification with ${\rm FLAG}=1$ only indicates that the received triple is or was generated by participant $i$. However, this alone does not tell whether the received triple is the current one. Indeed, an attacker could make use of this fact to launch two attacks that cannot be detected. First, the attacker could replace the current triple in $(i,j)$ by a previously observed triple that had been sent over $(i,j)$. Second, if there are multiple triples in $(i,j)$ simultaneously, the attacker could swap their orders in the link. In these two attacks, since the replaced or reordered triple is a valid triple of message and signatures, it can pass the Paillier verification algorithm ${\rm Alg_{ver}}$ and the third party cannot detect the attacks. However, with the index $\ell$, these two attacks cannot pass the verification operation in Algorithm \ref{algo:attack detection}, as a replaced or reordered triple does not match the index.





To proceed, we illustrate how to integrate Algorithm \ref{algo:attack detection} into Algorithm \ref{algo:privacy preserving TES}. At step 1, each participant $i\in\mathcal{V}\cup\{{\rm TP},{\rm CO}\}$ first generates a set of Paillier keys $(\alpha_i,\beta_i,\nu_i,\pi_i)$ by ${\rm Alg_{key}}$, broadcasts $(\alpha_i,\beta_i)$ and keeps $(\nu_i,\pi_i)$ private to itself. All these key generation operations are only performed once. Without loss of generality, we assume that all the public keys are sent over authenticated links enabled by a public-key infrastructure (PKI) \cite{CP-JP:2010}. Between step 2 and step 3, insert a step so that 
supplier $i\in\mathcal{V}_s$ (resp. customer $i\in\mathcal{V}_d$) as well as the third party runs $({\rm FLAG}_{i\ell}^s,\bar y_{i\ell}^s)={\rm Alg_{ad}}(i,{\rm TP},\alpha_i,\beta_i,\nu_i,\pi_i,y_{i\ell}^s,\ell)$ (resp. $({\rm FLAG}_{i\ell}^d,\bar y_{i\ell}^d)={\rm Alg_{ad}}(i,{\rm TP},\alpha_i,\beta_i,\nu_i,\pi_i,y_{i\ell}^d,\ell)$).
If ${\rm FLAG}_{i\ell}^s=1$ (resp. ${\rm FLAG}_{i\ell}^d=1$), then the third party adopts $\bar y_{i\ell}^s$ (resp. $\bar y_{i\ell}^d$) as $y_{i\ell}^s$ (resp. $y_{i\ell}^d$) at step 3. Then, between step 3 and step 4, insert another step so that
the third party as well as the coordinator runs $({\rm FLAG}_{\ell}^s,\bar y_{\ell}^s)={\rm Alg_{ad}}({\rm TP},{\rm CO},\alpha_{\rm TP},\beta_{\rm TP},\nu_{\rm TP},\pi_{\rm TP},y_{\ell}^s,2(\ell-1)+1)$ (resp. $({\rm FLAG}_{\ell}^d,\bar y_{\ell}^d)={\rm Alg_{ad}}({\rm TP},{\rm CO},\alpha_{\rm TP},\beta_{\rm TP},\nu_{\rm TP},\pi_{\rm TP},$ $y_{\ell}^d,2(\ell-1)+2)$). If ${\rm FLAG}_{\ell}^s=1$ (resp. ${\rm FLAG}_{\ell}^d=1$), then the coordinator adopts $\bar y_{\ell}^s$ (resp. $\bar y_{\ell}^d$) as $y_{\ell}^s$ (resp. $y_{\ell}^d$) at step 4. After $\lambda^*$ is derived at step 5, the coordinator and each agent $i\in\mathcal{V}$ run $({\rm FLAG}_{i},\bar\lambda_{i}^*)=({\rm CO},i,\alpha_{\rm CO},\beta_{\rm CO},\nu_{\rm CO},\pi_{\rm CO},10^{\sigma_{\lambda}}\lambda^*,i)$, where $\sigma_{\lambda}\in\mathbb{N}$ is the precision level of price, i.e., for any price $\lambda$, only the first $\sigma_{\lambda}$ decimal fraction digits are kept, while the rest are dropped. Hence, $10^{\sigma_{\lambda}}\lambda^*$ is a non-negative integer. If ${\rm FLAG}_{i}=1$, then agent $i$ uses $\bar\lambda_{i}^*$ as $\lambda^*$.

The above attack detection mechanism guarantees that any data injection attack can be detected by legitimate message receivers. This property directly follows from the security of the Paillier digital signature scheme and the usage of index.

\section{Case studies\label{sec:simulation}}

In this section, the proposed cyber-resilient design is tested on a TES that coordinates and controls residential air conditioners to manage the feeder congestion.


We consider the real-time electricity allocation of a distribution feeder on a hot summer day (August 16, 2009) for Columbus, Ohio, USA. The weather data and the Typical Meteorological Year (TMY2) data are adopted from \cite{Weather} and \cite{WM-KU:1995}. The wholesale resource price is adopted from the PJM market \cite{PJM} and it is modified to a retail rate plus a retail modifier as defined by American Electric Power (AEP)'s tariff \cite{AEP}. We define this retail price as the base price. The distribution feeder capacity limit is $3.5$ MW. There are 1000 residential ACs under the feeder. In this scenario, the feeder is both the coordinator and the only supplier, and each residential AC is a customer. In each market cycle, the feeder aims to obtain the aggregated demand curve and compares it with the feeder capacity limit to determine the market clearing price. If there is no congestion, then the clearing price is set to the base price. If there is congestion, the clearing price is set as the price corresponding to the feeder capacity limit on the aggregated demand curve. The price range is between $\lambda_{\min}=\$0$ and $\lambda_{\max}=\$1$ and the sampling period is $\tau=\$0.01$. We then have $N_p=101$. The length of a market cycle is 5 minutes. 
A second-order equivalent thermal parameter (ETP) model 
is used to capture the load dynamics of the ACs. Detailed description of the ETP model parameters can be found in \cite{GridLAB}.



\begin{figure}[!ht]
\begin{center}
\includegraphics[width=1\linewidth]{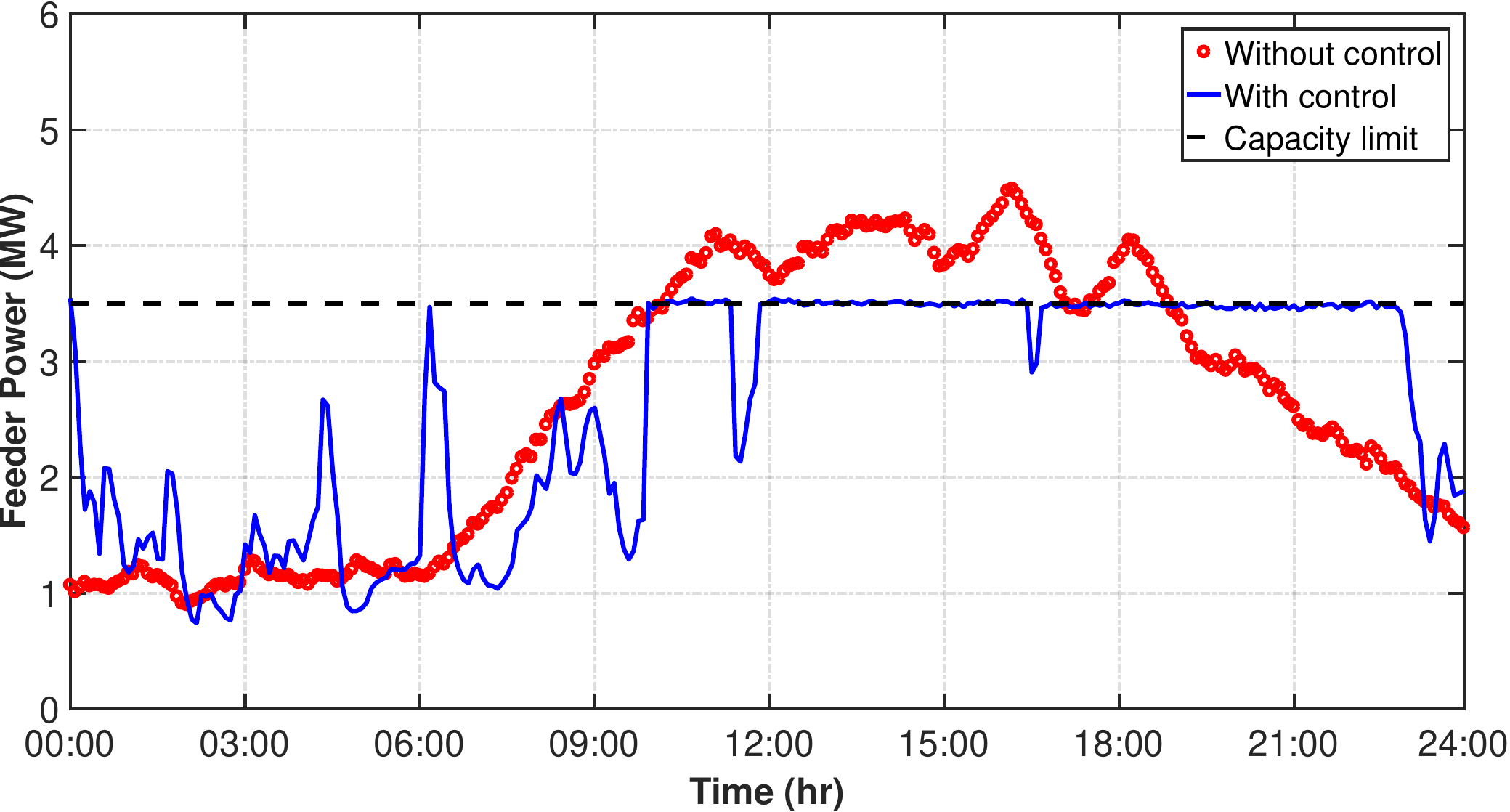}
\caption{Evolution of feeder power within 24 hours.}
\label{feeder_power}
\end{center}
\end{figure}

\begin{figure}[!ht]
\begin{center}
\includegraphics[width=1\linewidth]{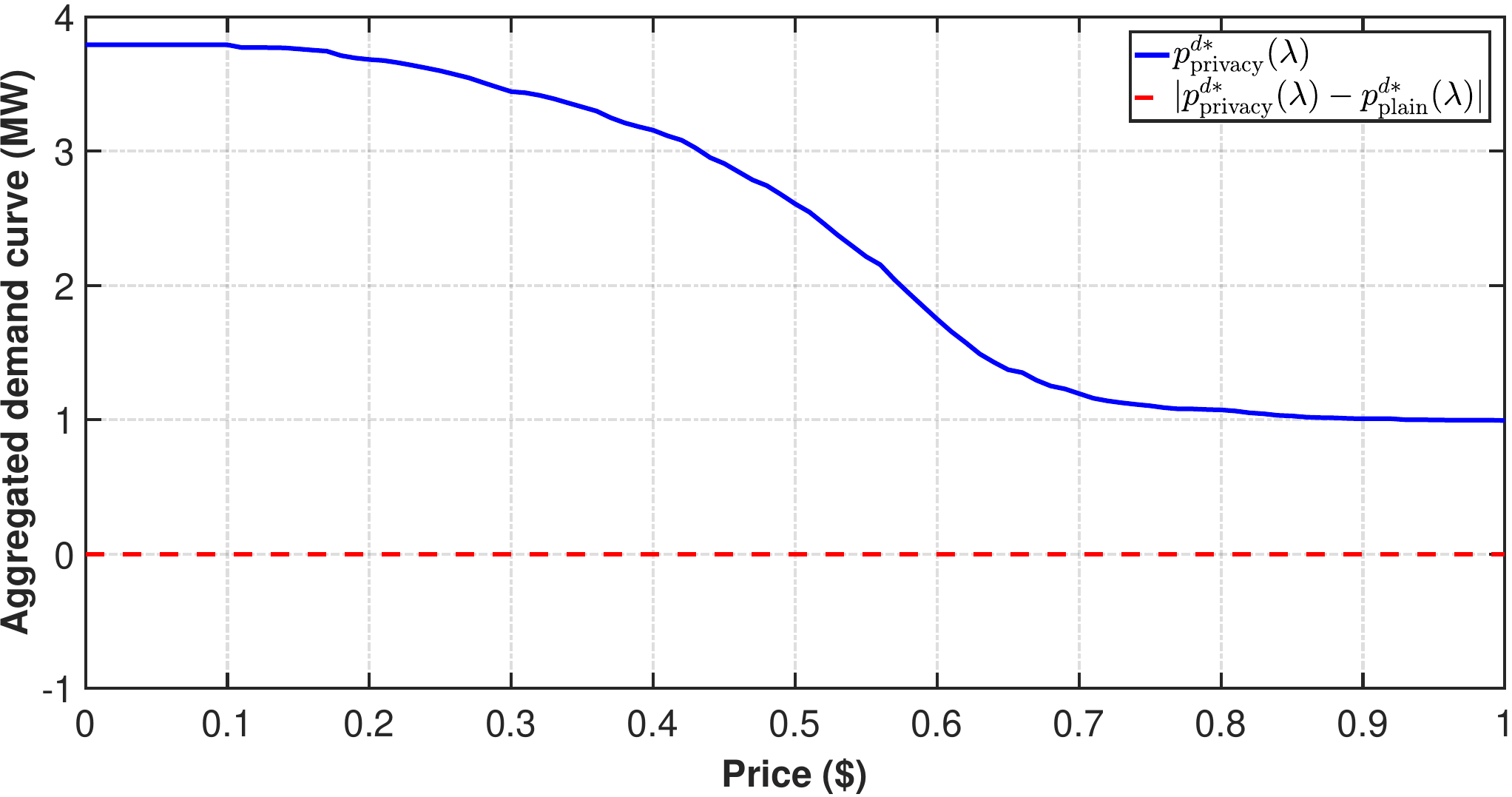}
\caption{Aggregated demand curve at 200-th market cycle.}
\label{demand_curve_simulation}
\end{center}
\end{figure}

\begin{figure}[!ht]
\begin{center}
\includegraphics[width=1\linewidth]{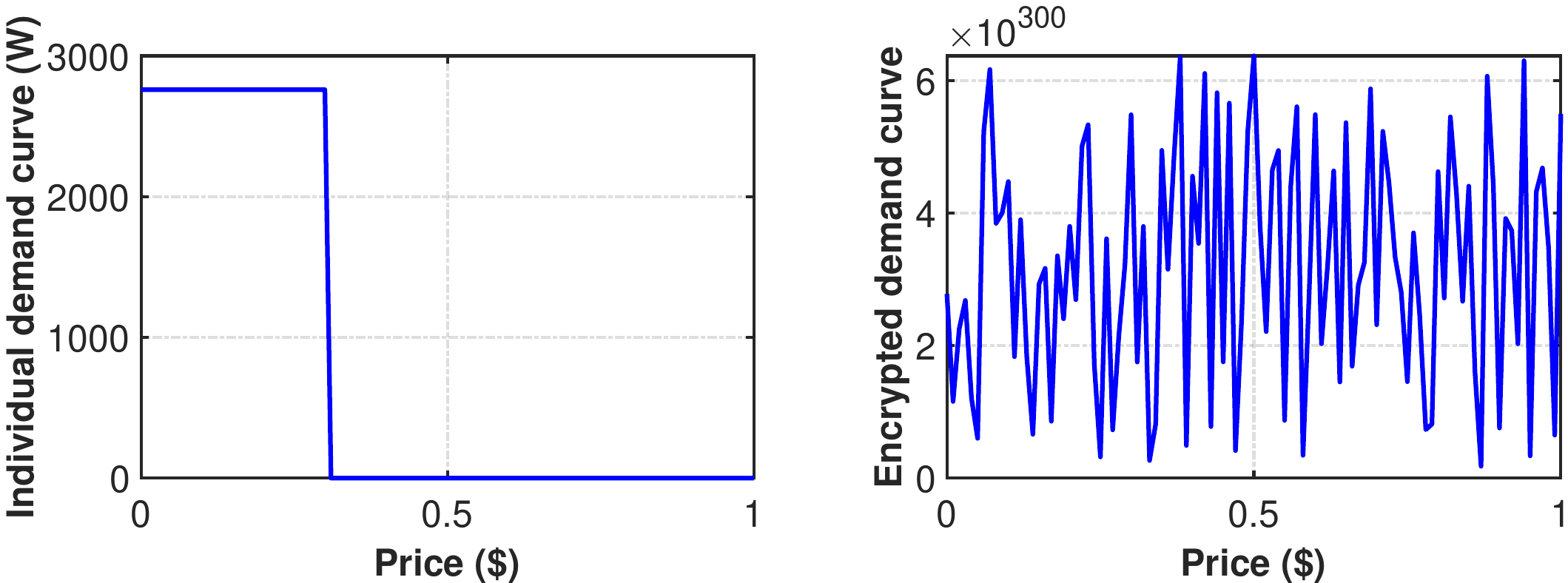}
\caption{Agent 100's demand curve at 200-th market cycle.}
\label{individual_demand_curve_simulation}
\end{center}
\end{figure}

We simulate the above problem for a whole day. We first verify the correctness and privacy preservation properties of Algorithm \ref{algo:privacy preserving TES} without data injection attacks. Fig \ref{feeder_power} shows the evolution of feeder power within 24 hours. The trajectory of feeder power with control (the solid blue line) is derived under the proposed privacy-preserving algorithm. Fig \ref{feeder_power} verifies that our algorithm maintains optimal market-based coordination. Fig. \ref{demand_curve_simulation} shows the aggregated demand curve at the 220-th market cycle (the number 220 is arbitrarily picked and any other market cycle can be used for illustration). Denote by $p^{d*}(\lambda)$ the aggregated demand curve, i.e., $p^{d*}(\lambda)\triangleq \sum_{i\in\mathcal{V}_d}p_i^{d*}(\lambda)$. We simulate the auction-based scheme both with and without our privacy-preserving design and denote the aggregated demand curves derived in the two cases by $p_{\rm privacy}^{d*}(\lambda)$ and $p_{\rm plain}^{d*}(\lambda)$, respectively. In Fig. \ref{demand_curve_simulation}, the curve $p_{\rm privacy}^{d*}(\lambda)$ (the solid blue line) shows the shape of the aggregated demand curve, and the curve $|p_{\rm privacy}^{d*}(\lambda)-p_{\rm plain}^{d*}(\lambda)|$ (the dashed red line), which is constant at 0, shows that $p_{\rm privacy}^{d*}(\lambda)$ is exactly equal to $p_{\rm plain}^{d*}(\lambda)$ at all values of $\lambda$, which verifies the correctness of Algorithm \ref{algo:privacy preserving TES}. In Fig. \ref{individual_demand_curve_simulation}, the left subfigure shows agent 100's demand curve at the 200-th market cycle, and the right subfigure shows its encryption under 500 bits of key length. Fig. \ref{individual_demand_curve_simulation} visually illustrates the privacy preservation of Algorithm \ref{algo:privacy preserving TES}, as the points of the encrypted demand curve look like pure random numbers within a large interval.

\begin{figure}[!ht]
\begin{center}
\includegraphics[width=1\linewidth]{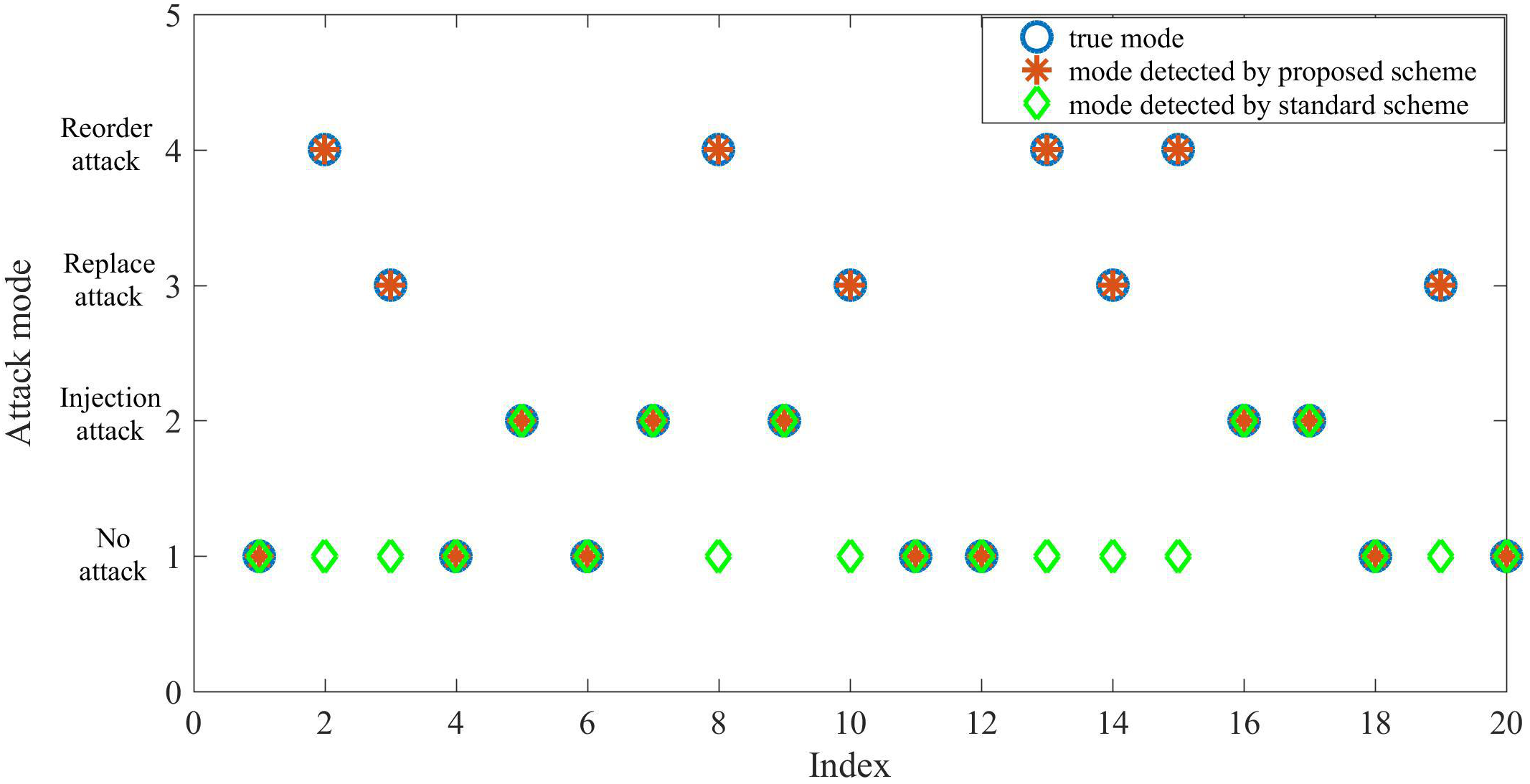}
\caption{Attack mode detection.}
\label{attack_mode_detection}
\end{center}
\end{figure}

Next we verify the security awareness of the attack detection mechanism in Section \ref{sec:address security issue}. We consider four different attack modes. Mode 1 is no attack. Mode 2 uses a randomly chosen message to replace the true message. More specifically, in this mode, to tamper a triple $(\ell\leftrightarrow m,s_1,s_2)$, an attacker randomly chooses a message $m'$ and generates a set of Paillier keys, and uses the keys to generate a pair of signatures $(s_1',s_2')$ for $\ell\leftrightarrow m'$. The triple $(\ell\leftrightarrow m',s_1',s_2')$ is sent to the message receiver. Mode 3 is data replace attack and mode 4 is data reorder attack. Please refer to Section \ref{sec:attack detection} for details of these two attacks. We deploy these four modes of attacks to agent 100's 20 consecutive messages at the 200-th market cycle. The detection result is shown in Fig. \ref{attack_mode_detection}. In the figure, the blue circle is the true attack mode, the red star is the attack mode detected by our proposed digital signature scheme, and the green diamond is the attack mode detected by the standard digital signature scheme in \cite{6578183}. Fig. \ref{attack_mode_detection} shows that our scheme is able to detect all the four modes of attacks, while the scheme in \cite{6578183} is only able to detect the attacks of modes 1 and 2.

Finally, we examine the efficiency of the integration of the proposed privacy-preserving and security-aware mechanisms. Table \ref{table:separate} lists the running time under different key lengths without and with the block design in Section \ref{sec:block design}. The time for agent (columns 2 and 5) is the average time per agent per market cycle, and the time for the third party and the coordinator (columns 3, 4, 6 and 7) is the average time per market cycle. We can see that, under the same key length, the running time with the block design is much smaller than that without the block design. For large key lengths, the rate between the running time without and with the block design is approximately $N_p=101$, which matches our expectation.

\begin{table}[!ht]
\renewcommand{\arraystretch}{1.2}
\caption{Comparison of Computational Overhead}
\centering
\begin{tabular}{|c|c|c|c|c|c|c|}
\hline
Key length & \multicolumn{3}{c|}{Point-wise (s)} & \multicolumn{3}{c|}{Block-wise (s)}\\
\cline{2-7}
(bit) & Agent & TP & CO & Agent & TP & CO\\
\hline
500 & 0.58 & 11.97& 0.53 & 0.016 & 0.12 & 0.010\\
\hline
1000 & 2.55 & 21.19 & 3.28 & 0.035 & 0.16 & 0.035\\
\hline
1500 & 7.61 & 26.57 & 10.02 & 0.089 & 0.27 & 0.11\\
\hline
2000 & 16.79 & 42.36 & 22.14 & 0.18 & 0.41 & 0.23\\
\hline
2500 & 32.70 & 70.03 & 42.79 & 0.34 & 0.71 & 0.44\\
\hline
3000 & 55.23 & 107.35 & 76.10 & 0.58 & 1.09 & 0.76\\
\hline
3500 & 87.51 & 157.78 & 114.63 & 0.88 & 1.61 & 1.20\\
\hline
4000 & 128.66 & 223.32 & 172.57 & 1.40 & 2.34 & 1.96\\
\hline
\end{tabular}
\label{table:separate}
\end{table}


\section{Conclusions\label{sec:conclusion}}

This paper studied the privacy and security issues associated with TESs. We first focused on the privacy issue by developing a homomorphic encryption-based algorithm to simultaneously achieve the optimal market-based coordination and privacy preservation. A block design was proposed to greatly improve the associated computational efficiency. Then, we moved to the security issue and proposed a digital signature-based mechanism that further ensures security awareness. 
The effectiveness of the proposed cyber-resilient TES design was verified by simulations on the transactive control of residential ACs.


\appendix{\label{sec:notation}}

The appendix serves to briefly introduce Paillier encryption and Paillier digital signature. More detailed discussions on Paillier cryptosystem can be found in \cite{PP:1999}.



\subsection{Paillier Encryption\label{sec:Paillier encryption}}

The Paillier encryption scheme is an additive homomorphic encryption scheme. It consists of key generation, encryption and decryption operations, as illustrated next.

$\bullet$ Key generation: A set of keys $(\alpha,\beta,\nu,\pi)$ is generated by Algorithm \ref{algo:key generation}, in which $n$ is the security parameter to set the key length,  $(\alpha,\beta)$ are public keys and broadcasted, while $(\nu,\pi)$ are private keys and kept secret to the executor itself.

\begin{algorithm2e}
\caption{Key generation algorithm}\label{algo:key generation}

\nonl Syntax: $(\alpha,\beta,\nu,\pi)={\rm Alg_{key}}(n)$.

\nonl The executor randomly chooses two large prime numbers $p$ and $q$ such that ${\rm gcd}(pq,(p-1)(q-1))=1$ and $|\alpha|=n$ with $\alpha=pq$; computes $\nu={\rm lcm}(p-1,q-1)$; randomly selects an integer $\beta\in\mathbb{Z}_{\alpha^2}^*$ such that the following modular multiplicative inverse $\pi$ exists
\[
\pi=\left(\frac{(\beta^\nu\mod\alpha^2)-1}{\alpha}\right)^{-1}\mod\alpha,
\]
i.e., $\pi\frac{(\beta^\nu\mod\alpha^2)-1}{\alpha}\equiv1\mod \alpha$.
\end{algorithm2e}

$\bullet$ Encryption: A plaintext $pt\in\mathbb{Z}_\alpha$ is encrypted as $ct$ with public keys $(\alpha,\beta)$ by Algorithm \ref{algo:encryption}.

\begin{algorithm2e}[htbp]
\caption{Encryption algorithm}\label{algo:encryption}

\nonl Syntax: $ct={\rm Alg_{enc}}(\alpha,\beta,pt)$.

\nonl The executor selects a random integer $r\in\mathbb{Z}_\alpha^*$ and computes $ct=\beta^{pt}\cdot r^\alpha\mod \alpha^2$.
\end{algorithm2e}

$\bullet$ Decryption: A ciphertext $ct\in\mathbb{Z}_{\alpha^2}$ is decrypted as $pt$ with public key $\alpha$ and private keys $(\nu,\pi)$ by Algorithm \ref{algo:decryption}.

\begin{algorithm2e}[htbp]
\caption{Decryption algorithm}\label{algo:decryption}

\nonl Syntax: $pt={\rm Alg_{dec}}(\alpha,\nu,\pi,ct)$.

\nonl The executor computes $pt=\frac{(ct^\nu \mod \alpha^2)-1}{\alpha}\cdot \pi \mod \alpha$.
\end{algorithm2e}


The correctness, privacy and homomorphic property of the Paillier encryption scheme are given as follows:

\noindent (i) Decryption correctness:
\[
{\rm Alg_{dec}}\left(\alpha,\nu,\pi,{\rm Alg_{enc}}(\alpha,\beta,pt)\right)=pt.
\]

\noindent (ii) Semantic security: If the decisional composite residuosity assumption (DCRA)\footnote{DCRA: Given a composite $C$ and an integer $z$, it is computationally intractable to decide whether $z$ is a $C$-residue modulo $C^2$ or not, i.e., whether there exists $y$ such that $z=y^C\mod C^2$.} holds, then the Paillier encryption scheme is semantically secure. That is, it is computationally infeasible for one to infer any information of plaintexts by observing the corresponding ciphertexts. In other words, this scheme does not disclose any information of plaintexts.

\noindent (iii) Homomorphic property: Given any $pt_1,\cdots,pt_m\in\mathbb{Z}_\alpha$. If $\sum_{\ell=1}^mpt_\ell\in\mathbb{Z}_\alpha$, then 
\[
{\rm Alg_{dec}}\left(\alpha,\nu,\pi,\prod_{\ell=1}^m{{\rm Alg_{enc}}}(\alpha,\beta,pt_\ell)\right)=\sum_{\ell=1}^m pt_\ell.
\]

%
%
%

\subsection{Paillier Digital Signature\label{sec:Paillier digital signature}}

The Paillier digital signature scheme consists of key generation, signature and verification operations, as illustrated next.
%
%
%

$\bullet$ Key generation: Same as the key generation operation of the Paillier encryption scheme.

$\bullet$ Signature: A pair of signatures $(s_1,s_2)$ is generated for a message $m\in\mathbb{Z}_{\alpha^2}$ with keys $(\alpha,\beta,\nu,\pi)$ by Algorithm \ref{algo:signing}.

\begin{algorithm2e}[htbp]
\caption{Signing algorithm}\label{algo:signing}

\nonl Syntax: $(s_1,s_2)={\rm Alg_{sig}}(\alpha,\beta,\nu,\pi,m)$.

\nonl The executor computes $s_1=\frac{(m^\nu\mod\alpha^2)-1}{\alpha}\cdot\pi\mod\alpha$ and $s_2=(m\cdot\beta^{-s_1})^{1/\alpha\mod\nu}\mod\alpha$.
\end{algorithm2e}

$\bullet$ Verification: A triple $(m,s_1,s_2)$ is verified with public keys $(\alpha,\beta)$ by Algorithm \ref{algo:verification}.

\begin{algorithm2e}[htbp]
\caption{Verification algorithm}\label{algo:verification}

\nonl Syntax: ${\rm FLAG}={\rm Alg_{ver}}(\alpha,\beta,m,s_1,s_2)$.

\nonl The executor sets ${\rm FLAG}=1$ if $m=\beta^{s_1}s_2^\alpha\mod\alpha^2$, and sets ${\rm FLAG}=0$ otherwise.
\end{algorithm2e}

The security of the Paillier digital signature is illustrated as follows: If the DCRA holds, then, after obtaining signatures to any messages of its choice, an attacker cannot generate a pair of signatures for a new message that can pass the verification with non-negligible probability.

\bibliographystyle{IEEEtran}
\bibliography{TES_CPS}

\begin{thebibliography}{10}
\providecommand{\url}[1]{#1}
\csname url@samestyle\endcsname
\providecommand{\newblock}{\relax}
\providecommand{\bibinfo}[2]{#2}
\providecommand{\BIBentrySTDinterwordspacing}{\spaceskip=0pt\relax}
\providecommand{\BIBentryALTinterwordstretchfactor}{4}
\providecommand{\BIBentryALTinterwordspacing}{\spaceskip=\fontdimen2\font plus
\BIBentryALTinterwordstretchfactor\fontdimen3\font minus
  \fontdimen4\font\relax}
\providecommand{\BIBforeignlanguage}[2]{{%
\expandafter\ifx\csname l@#1\endcsname\relax
\typeout{** WARNING: IEEEtran.bst: No hyphenation pattern has been}%
\typeout{** loaded for the language `#1'. Using the pattern for}%
\typeout{** the default language instead.}%
\else
\language=\csname l@#1\endcsname
\fi
#2}}
\providecommand{\BIBdecl}{\relax}
\BIBdecl

\bibitem{SL-JL-AC-WZ:2019}
S.~Li, J.~Lian, A.~Conejo, and W.~Zhang, ``Transactive energy system:
  {M}arket-based coordination of distributed energy resources,'' \emph{IEEE
  Control Systems Magazine}, August 2020.

\bibitem{YG-YC-YG-YF:2016}
Y.~Gong, Y.~Cai, Y.~Guo, and Y.~Fang, ``A privacy-preserving scheme for
  incentive-based demand response in the smart grid,'' \emph{IEEE Transactions
  on Smart Grid}, vol.~7, no.~3, pp. 1304--1313, 2016.

\bibitem{OT-DG-HVP:2013}
O.~Tan, D.~Gunduz, and H.~V. Poor, ``Increasing smart meter privacy through
  energy harvesting and storage devices,'' \emph{IEEE Journal on Selected Areas
  in Communications}, vol.~31, no.~7, pp. 1331--1341, 2013.

\bibitem{SH-UT-GJP:2016}
S.~Han, U.~Topcu, and G.~J. Pappas, ``Event-based information-theoretic
  privacy: A case study of smart meters,'' in \emph{Proc. of American Control
  Conference}, 2016, pp. 2074--2079.

\bibitem{ARB-DKM-BCL-PR:2013}
A.~R. Borden, D.~K. Molzahn, B.~C. Lesieutre, and P.~Ramanathan, ``Power system
  structure and confidentiality preserving transformation of optimal power flow
  problem,'' in \emph{Proc. Fifty-first Annual Allerton Conference}, 2013, pp.
  1021--1028.

\bibitem{ARB-DKM-PR-BCL:2012}
A.~R. Borden, D.~K. Molzahn, P.~Ramanathan, and B.~C. Lesieutre,
  ``Confidentiality-preserving optimal power flow for cloud computing,'' in
  \emph{Proc. Fiftieth Annual Allerton Conference}, 2012, pp. 1300--1307.

\bibitem{CD:06}
C.~Dwork, ``Differential privacy,'' in \emph{Proc. 3rd International Colloquium
  on Automata, Languages and Programming}, 2006, pp. 1--12.

\bibitem{Dwork}
C.~Dwork and A.~Roth, ``The algorithm foundations of differential privacy,''
  \emph{Foundations and Trends in Theoretical Computer Science}, vol.~9, no.
  3--4, pp. 211--407, August 2014.

\bibitem{ZY-PC-JC:2017}
Z.~Yang, P.~Cheng, and J.~Chen, ``Differential-privacy preserving optimal power
  flow in smart grid,'' \emph{IET Generation, Transmission and Distribution},
  vol.~11, no.~15, pp. 3853--3861, 2017.

\bibitem{FZ-JA-SHL:2019}
F.~Zhou, J.~Anderson, and S.~H. Low, ``Differential privacy of aggregated {DC}
  optimal power flow data,'' in \emph{Proc. American Control Conference}, 2019,
  pp. 1307--1314.

\bibitem{XL-RT-DKYY-PC:2017}
X.~Lou, R.~Tan, D.~K.~Y. Yau, and P.~Cheng, ``Cost of differential privacy in
  demand reporting for smart grid economic dispatch,'' in \emph{IEEE Conference
  on Computer Communications}, 2017, pp. 1--9.

\bibitem{7742416}
A.~{Halder}, X.~{Geng}, P.~R. {Kumar}, and L.~{Xie}, ``Architecture and
  algorithms for privacy preserving thermal inertial load management by a load
  serving entity,'' \emph{IEEE Transactions on Power Systems}, vol.~32, no.~4,
  pp. 3275--3286, 2017.

\bibitem{YL-MZ:2019ARC}
Y.~Lu and M.~Zhu, ``A control-theoretic perspective on cyber-physical privacy:
  {W}here data privacy meets dynamic systems,'' \emph{Annual Reviews in
  Control}, vol.~47, pp. 423--440, 2019.

\bibitem{DS-NH-GB:1995}
D.~Stevenson, N.~Hillery, and G.~Byrd, ``Secure communications in {ATM}
  networks,'' \emph{Communications of the ACM}, vol.~38, no.~2, pp. 45--52,
  1995.

\bibitem{LX-GRA:2001}
L.~Xie and G.~R. Arce, ``A class of authentication digital watermarks for
  secure multimedia communication,'' \emph{IEEE Transactions on Image
  Processing}, vol.~10, no.~11, pp. 1754--1764, 2001.

\bibitem{TJ-YH-SZ:2004}
T.~Jiang, Y.~Hou, and S.~Zheng, ``Secure communication between set-up box and
  smart card in {DTV} broadcasting,'' \emph{IEEE Transactions on Consumer
  Electronics}, vol.~50, no.~3, pp. 882--886, 2004.

\bibitem{6578183}
C.~{Fan}, S.~{Huang}, and Y.~{Lai}, ``Privacy-enhanced data aggregation scheme
  against internal attackers in smart grid,'' \emph{IEEE Transactions on
  Industrial Informatics}, vol.~10, no.~1, pp. 666--675, 2014.

\bibitem{PP:1999}
P.~Paillier, ``Public-key cryptosystems based on composite degree residuosity
  classes,'' in \emph{Proc. Advances in Cryptology, EUROCRYPT 1999}, 1999, pp.
  223--238.

\bibitem{YL-MZ:2015}
Y.~Lu and M.~Zhu, ``Secure cloud computing algorithms for discrete constrained
  potential games,'' \emph{Proc. 5th IFAC Workshop on Distributed Estimation
  and Control in Networked Systems}, vol.~48, no.~22, pp. 180--185, September
  2015.

\bibitem{YL-MZ:2018Automatica}
------, ``Privacy preserving distributed optimization using homomorphic
  encryption,'' \emph{Automatica}, vol.~96, no.~10, pp. 314--325, October 2018.

\bibitem{YS-KG-AA-GJP-SAS-MS-PT:2016}
Y.~Shoukry, K.~Gatsis, A.~Alanwar, G.~J. Pappas, S.~A. Seshia, M.~Srivastava,
  and P.~Tabuada, ``Privacy-aware quadratic optimization using partially
  homomorphic encryption,'' in \emph{Proc. 2016 IEEE 55th Conference on
  Decision and Control}, December 2016, pp. 5053--5058.

\bibitem{KK-TF:2015}
K.~Kogiso and T.~Fujita, ``Cyber-security enhancement of networked control
  systems using homomorphic encryption,'' in \emph{Proc. 54th IEEE Conference
  on Decision and Control}, December 2015, pp. 6836--6843.

\bibitem{FF-IS-NB:2017}
F.~Farokhi, I.~Shames, and N.~Batterham, ``Secure and private control using
  semi-homomorphic encryption,'' \emph{Control Engineering Practice}, vol.~67,
  pp. 13--20, October 2017.

\bibitem{NF-PP:2016}
N.~M. Freris and P.~Patrinos, ``Distributed computing over encrypted data,'' in
  \emph{Proc. 54th Annual Allerton Conference on Communication, Control, and
  Computing (Allerton)}, September 2016, pp. 1116--1122.

\bibitem{MR-HG-YW:2018}
M.~Ruan, H.~Gao, and Y.~Wang, ``Secure and privacy-preserving consensus,''
  \emph{IEEE Transactions on Automatic Control}, vol.~64, no.~10, pp.
  4035--4049, October 2019.

\bibitem{RP:2010}
R.~Petrlic, ``A privacy-preserving concept for smart grids,'' in \emph{Proc.
  Sicherheit in Vernetzten Systemen}, 2010, pp. B1--B14.

\bibitem{FDG-BJ:2010}
F.~D. Garcia and B.~Jacobs, ``Privacy-friendly energy-metering via homomorphic
  encryption,'' in \emph{Proc. International Workshop on Security and Trust
  Management}, 2010, pp. 226--238.

\bibitem{FL-BL-PL:2010}
F.~Li, B.~Luo, and P.~Liu, ``Secure information aggregation for smart grids
  using homomorphic encryption,'' in \emph{Proc. 1st IEEE International
  Conference on Smart Grid Communications}, 2010, pp. 327--332.

\bibitem{9444341}
T.~Wu, C.~Zhao, and Y.~Zhang, ``Privacy-preserving distributed optimal power
  flow with partially homomorphic encryption,'' \emph{IEEE Transactions on
  Smart Grid}, 2021, accepted.

\bibitem{YL-JL-MZ:2020}
Y.~Lu, J.~Lian, and M.~Zhu, ``Privacy-preserving transactive energy systems,''
  in \emph{Proc. American Control Conference}, 2020, pp. 3005--3010.

\bibitem{JL-HR-YS-DJH:2019}
J.~Lian, H.~Ren, Y.~Sun, and D.~J. Hammerstrom, ``Performance evaluation for
  transactive energy systems using double-auction market,'' \emph{IEEE
  Transactions on Power Systems}, vol.~34, no.~5, pp. 4128--4137, September
  2019.

\bibitem{IBMcloud}
\BIBentryALTinterwordspacing
R.~Gordon, \emph{{Power Systems in the IBM Cloud -- IBM Enterprise Level Cloud
  Support}}. [Online]. Available:
  \url{https://mainline.com/power-systems-in-the-ibm-cloud-enterprise-level-cloud-support/}
\BIBentrySTDinterwordspacing

\bibitem{Hazay}
C.~Hazay and Y.~Lindell, \emph{Efficient Secure Two-Party Protocols--Techniques
  and Constructions}.\hskip 1em plus 0.5em minus 0.4em\relax New York, NY:
  Springer, 2010.

\bibitem{DG:2017}
\BIBentryALTinterwordspacing
D.~Giry, ``Cryptographic key length recommendation,'' BlueKrypt, Tech. Rep.,
  2017. [Online]. Available: \url{https://www.keylength.com/en/8/}
\BIBentrySTDinterwordspacing

\bibitem{CP-JP:2010}
C.~Paar and J.~Pelzl, \emph{Understanding Cryptography}.\hskip 1em plus 0.5em
  minus 0.4em\relax Springer, 2010.

\bibitem{Weather}
\BIBentryALTinterwordspacing
\emph{Weather Uniderground: weather record for Columbus}. [Online]. Available:
  \url{https://www.wunderground.com/}
\BIBentrySTDinterwordspacing

\bibitem{WM-KU:1995}
W.~Marion and K.~Urban, ``User's manual for {TMY2s}: {T}ypical meteorological
  years: {D}erived from the 1961--1990 national solar radiation data base,''
  National Renewable Energy Lab, Golden, CO, Tech. Rep., 1995.

\bibitem{PJM}
\BIBentryALTinterwordspacing
\emph{{PJM} wholesale market energy price}. [Online]. Available:
  \url{http://pjm.com/markets-and-operations/energy.aspx}
\BIBentrySTDinterwordspacing

\bibitem{AEP}
\BIBentryALTinterwordspacing
\emph{{AEP} {O}hio power company standard tariff}. [Online]. Available:
  \url{https://aepohio.com/account/bills/rates/AEPOhioRatesTariffsOH.aspx}
\BIBentrySTDinterwordspacing

\bibitem{GridLAB}
\BIBentryALTinterwordspacing
\emph{{GridLAB-D} Residential Module User's Guide}. [Online]. Available:
  \url{http://www.eps.ee.kth.se/personal/luigiv/pst/}
\BIBentrySTDinterwordspacing

\end{thebibliography}

\begin{IEEEbiography}[{\includegraphics[width=1in,height=1.25in,clip,keepaspectratio]{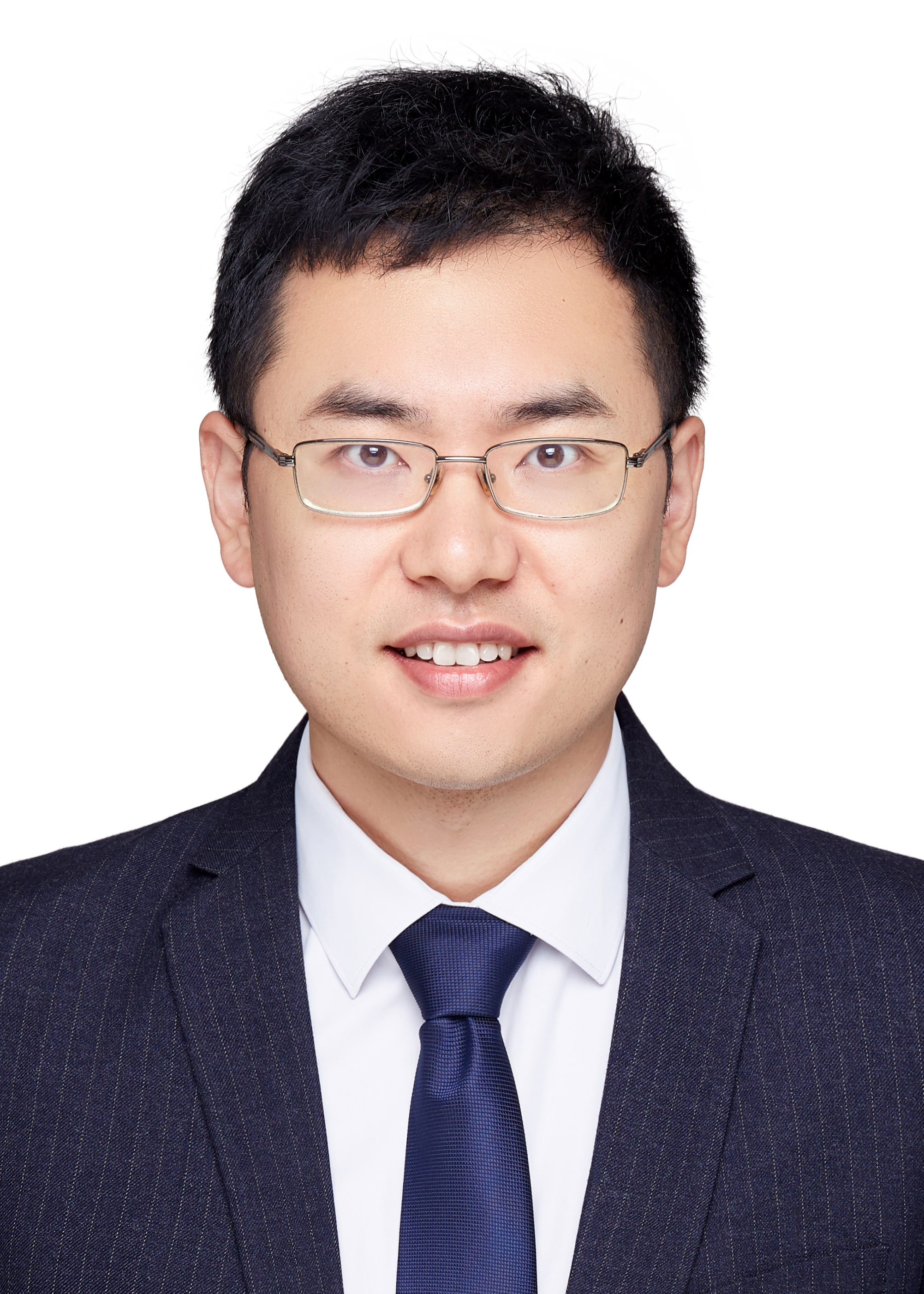}}]{Yang Lu}
is a Lecturer (Assistant Professor) of the Systems Security Group in the School of Computing and Communications at the Lancaster University. 
He received Ph.D. degree in Electrical Engineering from the Pennsylvania State University (PSU) in 2020, B.E. and M.E. degrees in Electrical Engineering from Shanghai Jiao Tong University in 2010 and 2013, respectively, and M.S. degree in Electrical Engineering from the Georgia Institute of Technology, in 2013. From September 2020 to August 2021, he worked as a postdoctoral scholar in the School of Electrical Engineering and Computer Science at PSU. From January 2019 to May 2019, he worked as a Ph.D. intern at the Pacific Northwest National Laboratory. From March 2013 to June 2014, he worked as a visiting scholar in the School of Electrical and Computer Engineering at the Georgia Institute of Technology. 
His research interests mainly focus on cyber-physical privacy and security, distributed control and optimization of multi-agent networks, and machine learning. He is a recipient of the Dr. Nirmal K. Bose Dissertation Excellence Award at PSU in 2019.
\end{IEEEbiography}

\begin{IEEEbiography}[{\includegraphics[width=1in,height=1.25in,clip,keepaspectratio]{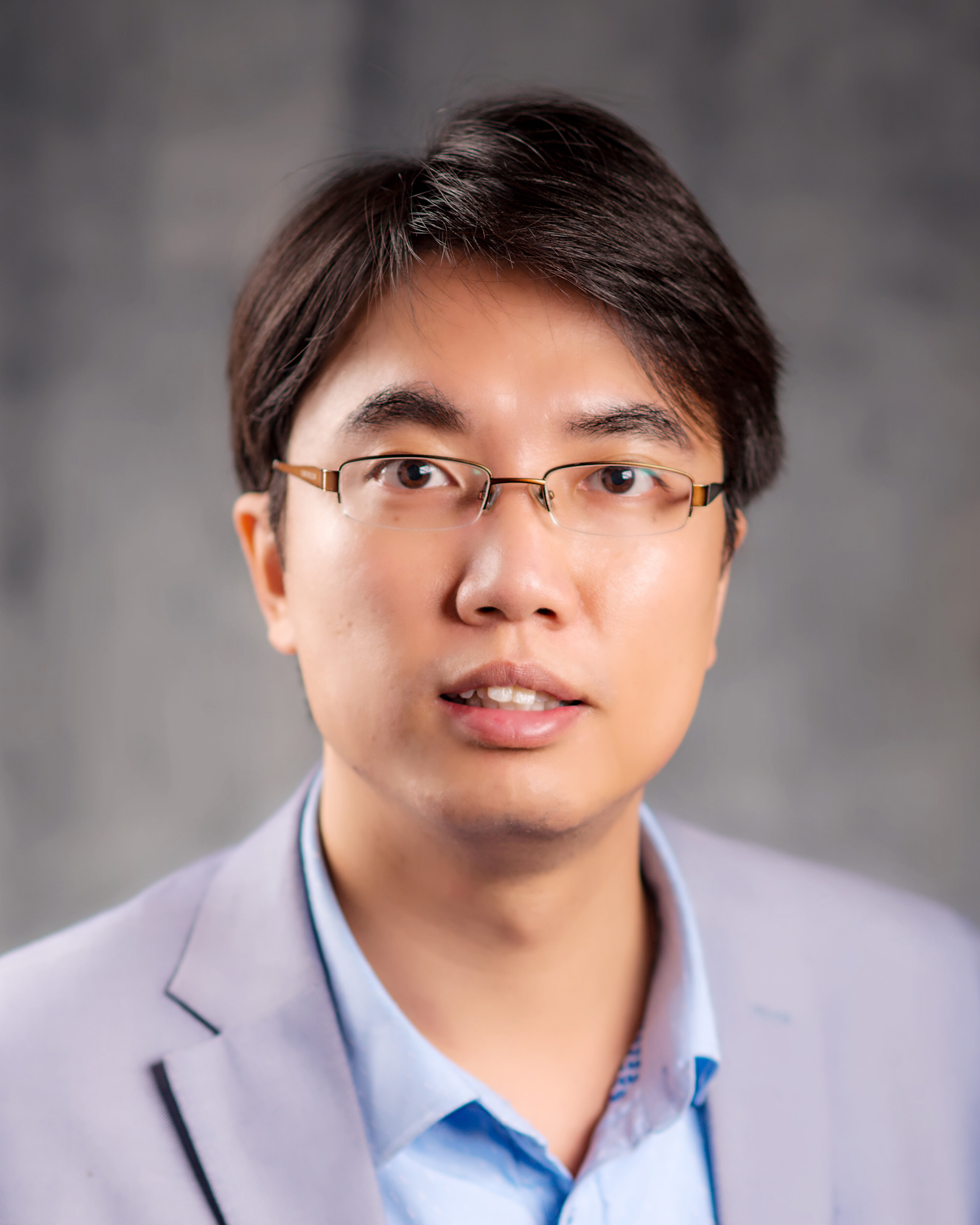}}]{Jianming Lian}
(S'09--M'10--SM'21) received the B.S. degree with the highest honor from the University of Science and Technology of China, Hefei, China, in 2004, and the M.S. and Ph.D. degrees in electrical engineering from Purdue University, West Lafayette, IN, USA, in 2007 and 2009, respectively. He is now a Distinguished R\&D staff and the group leader of Grid-interactive Controls Group in Energy Science and Technology Directorate at Oak Ridge National Laboratory (ORNL). Prior to that, he was a Chief Engineer and Team Lead in Energy and Environment Directorate at Pacific Northwest National Laboratory. He has served as the project manager, PI/Co-PI and key technical contributor of many large projects focusing on the engagement and integration of various distributed energy resources (DERs) into the future distribution management system. He has established the theoretical foundation of market-based control (aka. transactive control) for future transactive energy system. His research interests focus on the diverse methods from control, optimization, economics, game theory, data analytics and machine learning to improve the reliability and resilience as well as security and sustainability of complex energy systems including power grid and building system.
\end{IEEEbiography}

\begin{IEEEbiography}[{\includegraphics[width=1in,height=1.25in,clip,keepaspectratio]{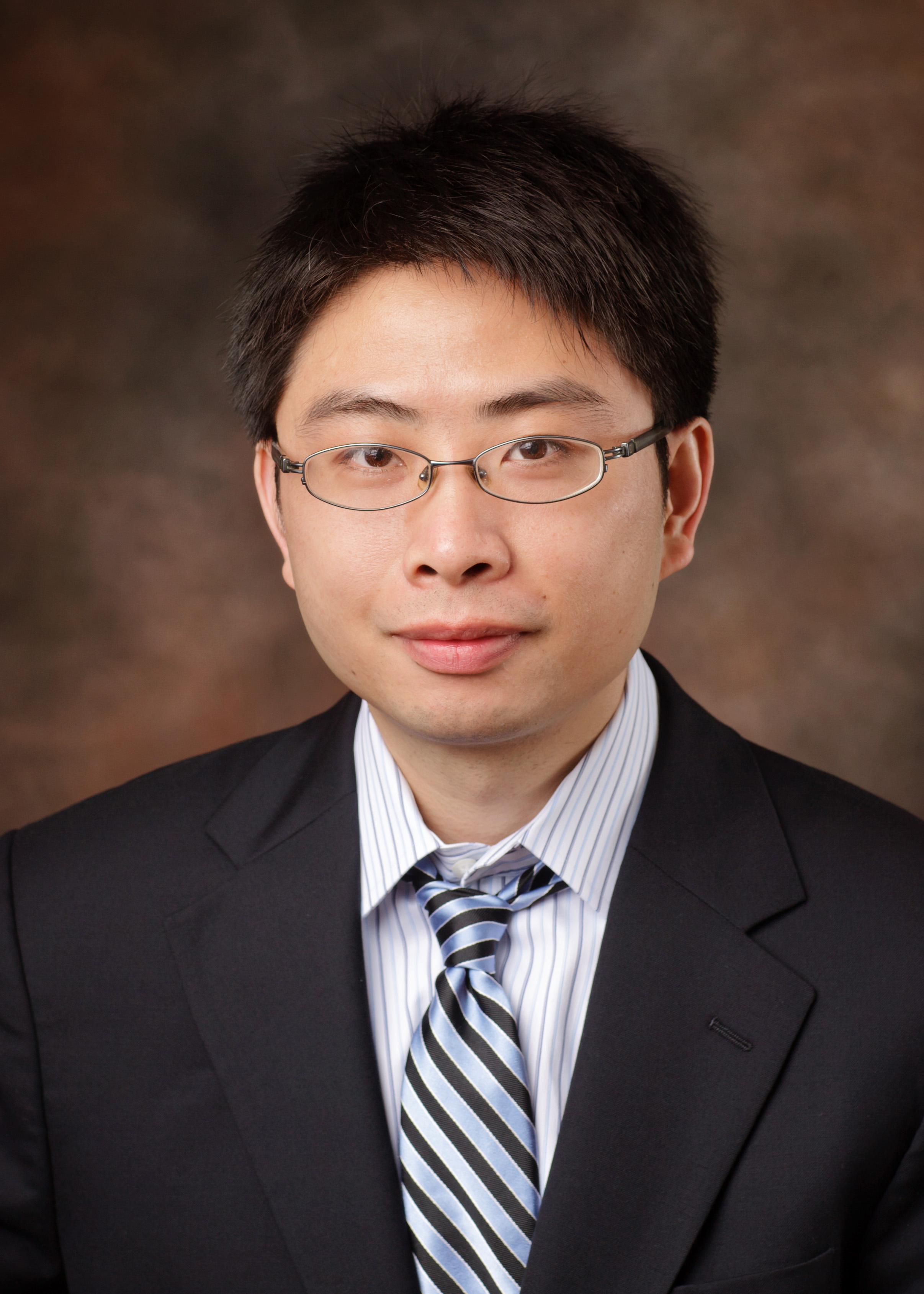}}]{Minghui Zhu}
is an Associate Professor in the School of Electrical Engineering and Computer Science at the Pennsylvania State University. Prior to joining Penn State in 2013, he was a postdoctoral associate in the Laboratory for Information and Decision Systems at the Massachusetts Institute of Technology. He received Ph.D. in Engineering Science (Mechanical Engineering) from the University of California, San Diego in 2011. His research interests lie in distributed control and decision-making of multi-agent networks with applications in robotic networks, security and the smart grid. He is the co-author of the book "Distributed optimization-based control of multi-agent networks in complex environments" (Springer, 2015). He is a recipient of the Dorothy Quiggle Career Development Professorship in Engineering at Penn State in 2013, the award of Outstanding Reviewer of Automatica in 2013 and 2014, and the National Science Foundation CAREER award in 2019. He is an associate editor of the IEEE Open Journal of Control Systems, the IET Cyber-systems and Robotics and the Conference Editorial Board of the IEEE Control Systems Society.
\end{IEEEbiography}

\begin{IEEEbiography}[{\includegraphics[width=1in,height=1.25in,clip,keepaspectratio]{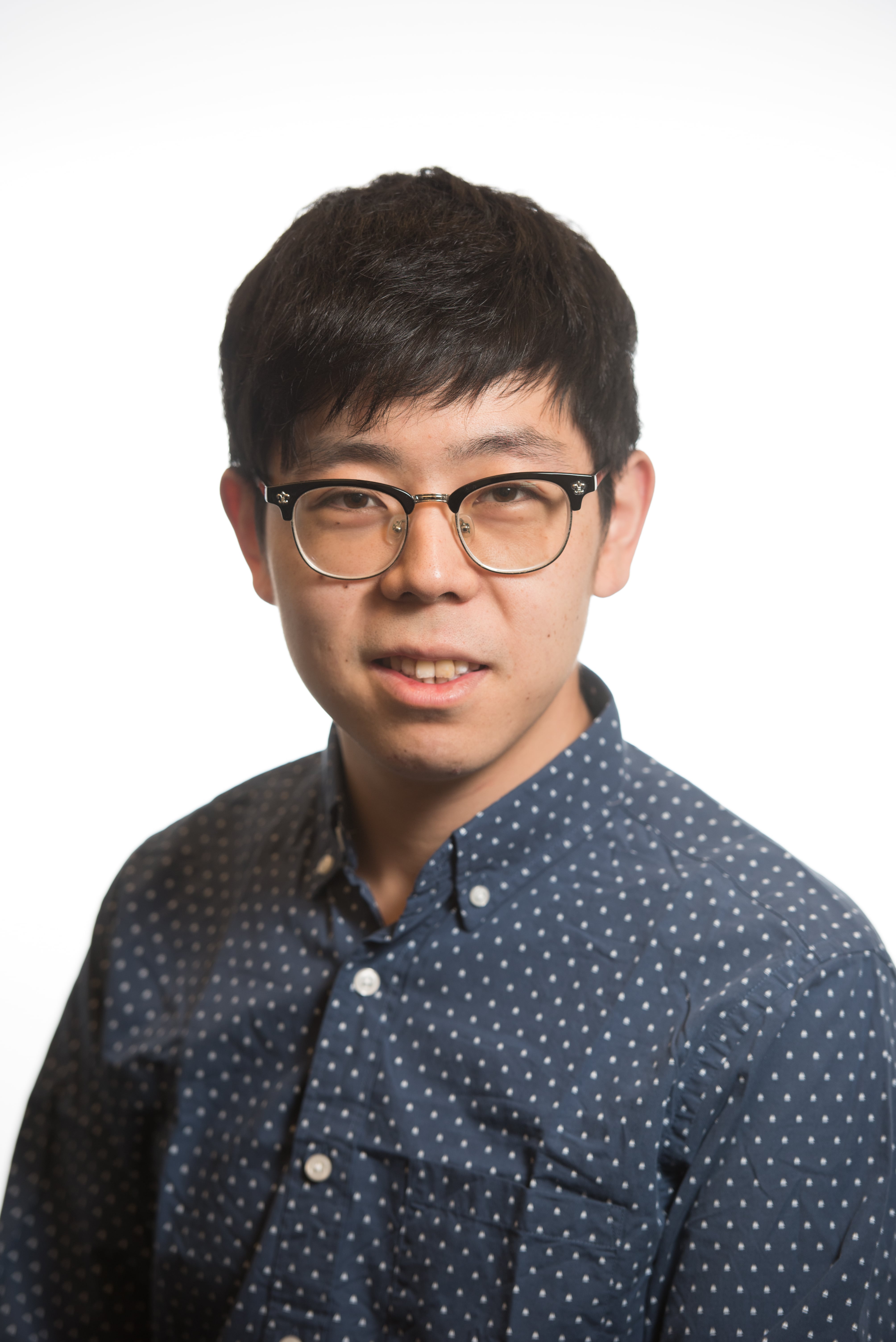}}]{Ke Ma}
received the B.E. degree in automation from Tsinghua University, Beijing, China, in 2012, and the Ph.D. degree in electrical and computer engineering from the Department of Electrical and Computer Engineering, Texas A\&M University, College Station, TX, USA in 2018. He is currently an electrical engineer at the Optimization and Control Group, Pacific Northwest National Laboratory (PNNL), Richland, WA, USA. His research interests include dynamic mechanism design and its application in electricity market, and market-based (transactive) coordination and control of distributed energy resources (DERs).
\end{IEEEbiography}

\end{document}